\documentclass[preprint,12pt,number]{elsarticle}

\usepackage{amsmath,amssymb}
\usepackage{graphicx}
\usepackage{booktabs}
\usepackage{algorithm}
\usepackage{algpseudocode}
\usepackage{hyperref}
\usepackage{xcolor}
\usepackage{subcaption}
\usepackage[acronym,nomain,nowarn]{glossaries}
\let\ac\gls   \let\Ac\Gls
\let\acs\acrshort  

\newacronym{tfce}{TFCE}{threshold-free cluster enhancement}
\newacronym{ptfce}{pTFCE}{probabilistic TFCE}
\newacronym{etfce}{eTFCE}{exact TFCE}
\newacronym{grf}{GRF}{Gaussian random field}
\newacronym{ccl}{CCL}{connected-component labelling}
\newacronym{fwer}{FWER}{family-wise error rate}
\newacronym{fdr}{FDR}{false discovery rate}
\newacronym{fwhm}{FWHM}{full width at half maximum}
\newacronym{vbm}{VBM}{voxel-based morphometry, used here to denote voxelwise analysis of spatially normalised T1-weighted intensity}
\newacronym{bet}{BET}{brain extraction tool}
\newacronym{mni}{MNI}{Montreal Neurological Institute}
\newacronym{roc}{ROC}{receiver operating characteristic}
\newacronym{qc}{QC}{quality control}

\graphicspath{{images/}}

\journal{NeuroImage}

\begin{document}

\begin{frontmatter}

\title{Hybrid eTFCE--GRF: Exact Cluster-Size Retrieval with Analytical $p$-Values for Voxel-Based Morphometry}

\author[cam]{Don Yin\corref{cor}\fnref{orcid1}}
\ead{dy323@cam.ac.uk}
\author[cam]{Hao Chen\corref{cor}\fnref{orcid2}}
\ead{hc666@cam.ac.uk}
\author[cam_eng]{Takeshi Miki\fnref{orcid5}}
\ead{tm858@cam.ac.uk}
\author[nus]{Enyu Yang\fnref{orcid4}}
\ead{yey@u.nus.edu}

\cortext[cor]{Corresponding author.}
\fntext[orcid1]{ORCID: \href{https://orcid.org/0000-0002-8971-1057}{0000-0002-8971-1057}}
\fntext[orcid4]{ORCID: \href{https://orcid.org/0000-0003-1441-1276}{0000-0003-1441-1276}}
\fntext[orcid2]{ORCID: \href{https://orcid.org/0009-0009-5281-4616}{0009-0009-5281-4616}}
\fntext[orcid5]{ORCID: \href{https://orcid.org/0009-0000-2063-6274}{0009-0000-2063-6274}}
\affiliation[cam]{organization={Department of Clinical Neurosciences, University of Cambridge},
                   city={Cambridge},
                   country={United Kingdom}}
\affiliation[cam_eng]{organization={Department of Engineering, University of Cambridge},
                   city={Cambridge},
                   country={United Kingdom}}
\affiliation[nus]{organization={National University of Singapore},
                   city={Singapore},
                   postcode={117583},
                   country={Singapore}}

\begin{highlights}
\item Hybrid eTFCE--GRF: exact cluster-size retrieval with analytical inference
\item FWER controlled at nominal level in Monte Carlo validation
\item \texttt{pytfce} package 4.6$\times$ (hybrid) to 75$\times$ (baseline) faster than R pTFCE
\item Validated on UK Biobank ($N = 500$) and IXI ($N = 563$)
\item Open-source Python package: \texttt{pip install pytfce}
\end{highlights}

\begin{abstract}
Threshold-free cluster enhancement (TFCE) improves sensitivity in voxel-wise neuroimaging inference by integrating cluster extent across all thresholds, but its reliance on permutation testing makes it prohibitively slow for large datasets.
Probabilistic TFCE (pTFCE) replaces permutations with analytical Gaussian random field (GRF) $p$-values, which reduces runtime by more than an order of magnitude, yet relies on a fixed threshold grid that introduces discretisation error.
Exact TFCE (eTFCE) eliminates this discretisation by computing the integral exactly via a union-find data structure, but still requires permutations for inference.
We propose a hybrid method that combines eTFCE's union-find data structure for exact cluster-size retrieval with pTFCE's analytical GRF inference.
The union-find builds the full cluster hierarchy in a single pass over sorted voxels and enables exact cluster-size queries at any threshold in near-constant time; GRF theory then converts these sizes into analytical $p$-values without permutations.
We validate the method through a six-experiment Monte Carlo study on synthetic phantoms ($64^3$, 80 subjects): null family-wise error rate is controlled at the nominal level (0/200 rejections, 95\% CI $[0.0\%, 1.9\%]$); power curves match baseline pTFCE (Dice $\geq 0.999$ at sufficient signal); smoothness estimation error is below 1\%; and cross-variant concordance exceeds $r = 0.99$.
On real brain data from UK Biobank ($N = 500$, within-vendor) and IXI ($N = 563$, cross-vendor), the method detects biologically plausible scanner, age, and sex effects; on IXI, significance maps form strict subsets of the reference R~pTFCE output, which supports conservative family-wise error control.
Both methods are implemented in \texttt{pytfce}, a pure-Python package with no R or FSL dependencies, available on PyPI.
The baseline reimplementation completes whole-brain voxel-based morphometry in ${\sim}5$\,s ($75\times$ faster than R~pTFCE), while the hybrid variant completes in ${\sim}85$\,s ($4.6\times$ faster) with the advantage of exact cluster-size retrieval; both are more than three orders of magnitude faster than permutation-based TFCE.
\end{abstract}


\end{frontmatter}

\glsresetall

\section{Introduction}
\label{sec:introduction}

Cluster-based thresholding in neuroimaging exploits spatial structure to boost sensitivity, but the choice of cluster-forming threshold strongly influences results, particularly at lower thresholds \cite{smith2009,friston1994}.
\Ac{tfce} eliminates this dependence by integrating cluster extent across all threshold levels, weighted by powers of height and spatial extent \cite{smith2009}.
Because the resulting scores lack a known null distribution, $p$-values in practice require permutation testing, which demands thousands of relabellings and takes hours to days on whole-brain \ac{vbm} data \cite{ashburner2000}.
This computational burden is particularly acute for biobank-scale studies such as the UK Biobank \cite{miller2016}, where thousands of statistical maps must be corrected for multiple comparisons.

Two recent methods address complementary limitations of \ac{tfce}.
\Ac{ptfce} replaces permutations with analytical $p$-values derived from \ac{grf} theory, which reduces runtime by more than an order of magnitude \cite{spisak2019}.
However, \ac{ptfce} evaluates cluster sizes by \ac{ccl} at a fixed grid of thresholds (typically 100), and the resulting discretisation introduces approximation error that depends on grid spacing.
\Ac{etfce} eliminates this discretisation by exploiting the piecewise-constant structure of the cluster-extent function to evaluate the \ac{tfce} integral in closed form, with a union-find data structure for efficient cluster retrieval \cite{chen2026}.
Chen et al.\ also identified a long-standing scaling error in FSL's \ac{tfce} implementation (the step size $\Delta\tau$ is omitted from the discrete approximation, confirmed up to version 6.0.7.19).
Although \ac{etfce} halves computation time relative to FSL's implementation, it retains the permutation requirement for inference, so that whole-brain analyses still take hours to days.

Neither method achieves both properties simultaneously: \ac{ptfce} is fast but approximate; \ac{etfce} is exact but slow.
This is an unnecessary trade-off: the two advances are algorithmically complementary, and there is no technical barrier to combining them.
We propose a hybrid \ac{etfce}--\ac{grf} method that does exactly this: it combines the union-find data structure of \ac{etfce} for exact cluster-size retrieval with the analytical \ac{grf} inference of \ac{ptfce}.
The union-find builds the full cluster hierarchy in a single pass over sorted voxels and enables exact cluster-size queries at any threshold in near-constant time; \ac{grf} theory then converts these sizes into analytical $p$-values without permutations.

We validate the method through a six-experiment Monte Carlo study and evaluate it on real brain data from UK Biobank ($N = 500$, within-vendor) and IXI ($N = 563$, cross-vendor).
The method is implemented in \texttt{pytfce}, a pure-Python package available on PyPI, with a companion software paper submitted to the Journal of Open Source Software \cite{yin2026}.

The contributions of this work are:
\begin{enumerate}
\item A hybrid algorithm that combines \ac{etfce}'s union-find data structure for exact cluster-size retrieval with \ac{ptfce}'s analytical \ac{grf} inference and achieves both properties simultaneously for the first time.
\item A six-experiment Monte Carlo validation that provides evidence of nominal \ac{fwer} control, no power loss relative to baseline \ac{ptfce}, accurate smoothness estimation, and high cross-variant concordance.
\item Evaluation on two real brain cohorts (UK Biobank $N = 500$ and IXI $N = 563$) that the method detects biologically plausible scanner, age, and sex effects, with runtime $4.6\times$ to $75\times$ faster than the reference R implementation, where the speedup depends on the variant.
\item An open-source Python package, \texttt{pytfce}, available on PyPI with no R or FSL dependencies.
\end{enumerate}

\section{Background}
\label{sec:background}

\subsection{Threshold-free cluster enhancement}
\label{sec:tfce}

Given a statistical map with voxel values $\{h_v\}$, \ac{tfce} assigns each voxel $v$ an enhanced score by integrating cluster extent across all threshold levels \cite{smith2009}:
\begin{equation}
\label{eq:tfce}
\mathrm{TFCE}(v) = \int_{h_0}^{h_v} e_v(h)^{E}\, h^{H}\,\mathrm{d}h,
\end{equation}
where $h_0 \geq 0$ is the minimum threshold (typically $h_0 = 0$), $e_v(h)$ is the number of voxels in the connected component containing $v$ when the image is thresholded at height $h$ (using 26-connectivity in 3D), and $E$ and $H$ are exponent parameters.
Setting $E = 0.5$ and $H = 2$ is standard: $H = 2$ reflects the quadratic relationship $-\log P(Z \geq h) \approx h^2/2$ under Gaussianity, while $E = 0.5$ was selected by \ac{roc} optimisation and is broadly consistent with the RFT-derived exponent $E = 2/3$ for the cluster-size survival function in three dimensions \cite{smith2009}.
\Ac{tfce} with $E = 1$, $H = 0$ reduces to the cluster mass statistic \cite{bullmore1999}.

In practice, the integral is approximated by a Riemann sum over $n$ equally spaced thresholds $\{\tau_i\}$ with step size $\Delta\tau = h_{\max}/n$, where $h_{\max} = \max_v h_v$:
\begin{equation}
\label{eq:tfce-discrete}
\mathrm{TFCE}(v) \approx \sum_{i=1}^{n} e_v(\tau_i)^{E}\, \tau_i^{H}\, \Delta\tau.
\end{equation}
FSL uses a fixed step size $\Delta\tau = 0.1$ \cite{smith2009}.
Each evaluation of $e_v(\tau_i)$ requires \ac{ccl} of the thresholded image, so the total cost scales as $O(nN)$ where $N$ is the number of voxels.
Because the \ac{tfce} score lacks a known null distribution, inference relies on permutation testing: for each of $B$ relabellings, the maximum \ac{tfce} score across all voxels is recorded, and the observed scores are compared against this null distribution to obtain \ac{fwer}-corrected $p$-values \cite{smith2009,winkler2014}.
For whole-brain analyses this requires thousands of relabellings, which makes \ac{tfce} computationally expensive for large cohorts.

\subsection{Probabilistic TFCE}
\label{sec:ptfce}

To avoid the computational cost of permutation testing, \ac{ptfce} derives voxel-level significance analytically from \ac{grf} theory \cite{spisak2019}.
At each threshold $\tau_i$ on a grid of $n$ levels (typically 100, equidistant in $-\log P$ space for greater accuracy at small $p$-values \cite{spisak2019}), \ac{ptfce} computes a conditional probability $P(Z_v \geq \tau_i \mid c_i)$ that combines voxel-level and cluster-level evidence via Bayes' theorem, where $c_i = e_v(\tau_i)$ is the observed cluster size.

The cluster-size distribution under the null is derived from \ac{grf} theory.
The expected number of clusters at threshold $h$ is approximated by the expected Euler characteristic \cite{worsley1996,friston1994}:
\begin{equation}
\label{eq:ec}
\mathrm{E}[\chi_h] = R_3\, (h^2 - 1)\, \exp(-h^2/2)\, (2\pi)^{-2},
\end{equation}
where $R_3 = N\,|\Lambda|^{1/2}$ is the three-dimensional RESEL count, $N$ is the number of voxels, and $|\Lambda|^{1/2}$ is the roughness of the field estimated from the data.
This approximation assumes that the field is sufficiently smooth (\ac{fwhm} exceeding three times the voxel size), approximately stationary, and that the threshold is high enough for the Euler characteristic to count isolated clusters rather than a single connected component \cite{worsley1996}.
The expected cluster size is $\mathrm{E}[c_h] = N(1 - \Phi(h)) / \mathrm{E}[\chi_h]$, where $\Phi$ is the standard normal CDF, and the cluster-size survival function follows from the result that $c^{2/3}$ is approximately exponentially distributed under \ac{grf} \cite{nosko1969}:
\begin{equation}
\label{eq:cluster-surv}
P(C > c \mid h) = \exp\!\Bigl(-\lambda_h\, c^{2/3}\Bigr), \quad \lambda_h = \Bigl(\mathrm{E}[c_h]\,/\,\Gamma\!\left(\tfrac{5}{2}\right)\Bigr)^{\!-2/3}.
\end{equation}

Because the conditional probabilities at successive thresholds form an incremental series rather than a pool of independent beliefs about the same event, the accumulated evidence $S(v)$ cannot be aggregated by a standard method such as Fisher's combination test.
Spis\'ak et al.\ introduced a correction function $Q$ that maps the sum of $-\log P$ scores to an equivalent single-test significance level \cite{spisak2019}:
\begin{equation}
\label{eq:ptfce-sum}
S(v) = \sum_{i=1}^{n} -\log P(Z_v \geq \tau_i \mid c_i),
\end{equation}
\begin{equation}
\label{eq:q-function}
\mathrm{pTFCE}(v) = Q\bigl(S(v)\bigr) = \frac{\sqrt{\Delta\,(8\,S(v) + \Delta)} - \Delta}{2},
\end{equation}
where $\Delta$ is the constant increment between successive thresholds in $-\log P$ space.
The output is in units of $-\log \bar{P}$ and can be corrected for multiple comparisons using standard thresholds from the unenhanced image (\ac{fdr} or Bonferroni), without permutation testing.
However, \ac{ptfce} retains the discretisation: cluster sizes and probabilities are evaluated at $n$ fixed thresholds, so accuracy depends on grid density.

\subsection{Exact TFCE}
\label{sec:etfce}

\Ac{etfce} eliminates discretisation error by exploiting the observation that $e_v(h)$ is piecewise constant: the cluster extent changes only at heights corresponding to actual voxel values \cite{chen2026}.
For each voxel $v$, the set of change points $\mathcal{T}_v$ defines intervals on which $e_v(h)$ is constant, and the \ac{tfce} integral evaluates in closed form:
\begin{equation}
\label{eq:etfce}
\mathrm{TFCE}(v) = \frac{1}{H+1} \sum_{i=1}^{|\mathcal{T}_v|} e_v(\tau_i)^{E} \Bigl(\tau_i^{H+1} - \tau_{i-1}^{H+1}\Bigr),
\end{equation}
where $\tau_0 = 0$ and $\tau_{|\mathcal{T}_v|} = h_v$.

The change points and cluster sizes are computed efficiently using a union-find data structure \cite{chen2026}.
Voxels are processed in descending order of their statistic values; each voxel is merged with any already-processed neighbours, which maintains a tree in which each connected component corresponds to a supra-threshold cluster.
The complexity is $O(N \log N)$, dominated by the initial sort, compared with $O(nN)$ for the Riemann-sum approximation.

Chen et al.\ showed that several major cluster-based statistics are special cases of a generalised formulation:
\begin{equation}
\label{eq:generalised}
T(v) = \int_{h_0}^{\infty} g\bigl(e_v(h)\bigr)\, f(h)\,\mathrm{d}h,
\end{equation}
of which $\mathrm{TFCE}(v)$ is a special case with $g(x) = x^E$, $f(h) = h^H$; the cluster mass statistic corresponds to $g(x) = x$, $f(h) = 1$ \cite{chen2026}.
The union-find framework computes multiple such statistics in a single pass at negligible additional cost.

Despite its exactness, \ac{etfce} retains the permutation requirement: $p$-values are obtained by comparing observed scores against a null distribution built from sign-flipped or permuted data.
The union-find already provides exact cluster sizes at arbitrary thresholds, which is precisely what \ac{grf}-based inference requires as input, yet no existing method combines exact cluster-size retrieval with analytical inference.

\subsection{Summary}
\label{sec:bg-summary}

Table~\ref{tab:comparison} summarises the properties of the three methods.
\Ac{ptfce} solves the speed problem (analytical inference, no permutations) but inherits the discretisation of the original \ac{tfce}.
\Ac{etfce} solves the discretisation problem (exact integral via union-find) but retains the permutation requirement.
The hybrid method proposed in Section~\ref{sec:methods} combines both advances.

\begin{table}[t]
\centering
\caption{Properties of existing cluster-enhancement methods.
\acs{tfce}~\cite{smith2009}, \acs{ptfce}~\cite{spisak2019}, and \acs{etfce}~\cite{chen2026} each address a different limitation; no single method achieves both exact cluster-size retrieval and analytical inference.
$B$ denotes the number of permutations, $n$ the number of threshold levels, and $N$ the number of voxels.}
\label{tab:comparison}
\footnotesize
\setlength{\tabcolsep}{4pt}
\begin{tabular}{@{} l p{2.6cm} p{2.6cm} p{2.6cm} @{}}
\toprule
 & \textbf{\acs{tfce}} & \textbf{\acs{ptfce}} & \textbf{\acs{etfce}} \\
\midrule
Integration
  & Riemann sum ($n$ thresholds)
  & Riemann sum ($n$ thresholds)
  & Exact (union-find) \\[3pt]
Inference
  & Permutation ($B$ relabellings)
  & \acs{grf} analytical
  & Permutation ($B$ relabellings) \\[3pt]
Output
  & Enhanced score (arb.\ units)
  & $-\log\bar{P}$ per voxel
  & Exact \acs{tfce} score \\[3pt]
Correction
  & \acs{fwer} only
  & \acs{fdr} or \acs{fwer}
  & \acs{fwer} only \\[3pt]
Assumptions
  & None (non-parametric)
  & \acs{grf} (stationarity, smoothness)
  & None (non-parametric) \\[3pt]
Complexity
  & $O(BnN)$
  & $O(nN)$
  & $O(BN\log N)$ \\
\bottomrule
\end{tabular}
\end{table}

\section{Methods}
\label{sec:methods}

\subsection{Hybrid eTFCE--GRF algorithm}
\label{sec:hybrid}

The two recent advances described in Section~\ref{sec:background} address complementary limitations of \ac{tfce}.
\Ac{ptfce} replaces permutation testing with analytical \ac{grf} inference and reduces runtime by more than an order of magnitude, but evaluates cluster sizes via \ac{ccl} at $n$ fixed thresholds, which introduces a discretisation error that increases with grid spacing \cite{spisak2019}.
\Ac{etfce} eliminates this discretisation by computing cluster sizes via a union-find data structure in a single descending sweep over sorted voxels, but retains the permutation requirement for inference \cite{chen2026}.
The hybrid method proposed here replaces \ac{ptfce}'s \ac{ccl}-based cluster retrieval with \ac{etfce}'s union-find and retains \ac{ptfce}'s \ac{grf}-based analytical inference, which achieves both exact cluster-size queries and permutation-free $p$-values in a single framework.
This combination is natural because the union-find is agnostic to how its cluster sizes are consumed: whether they feed into a permutation null (as in \ac{etfce}) or into an analytical \ac{grf} survival function (as here) is simply a downstream choice, not a change to the data structure itself.

The union-find data structure is constructed as follows.
All $N$ voxels are sorted by descending statistic value.
Voxels are then processed in this sorted order: for each voxel~$v$, (i) a singleton set $\{v\}$ is created, and (ii) for each already-processed neighbour~$u$ in the 26-connectivity neighbourhood, the sets containing $v$ and $u$ are merged using union-by-rank with path compression \cite{tarjan1975}.
After all voxels have been processed, the resulting disjoint-set forest encodes the complete merge tree of the superlevel-set filtration.
For any threshold~$h$, the cluster containing voxel~$v$ can be retrieved by finding the root of $v$'s component in the forest restricted to voxels with statistic $\geq h$.
The cluster size at that threshold is stored as an attribute of the root node.
Each find operation with path compression has worst-case amortised cost $O(\alpha(N))$, where $\alpha$ denotes the inverse Ackermann function \cite{tarjan1975}; $\alpha(N) \leq 5$ for all $N$ below $2^{65536}$, so each cluster-size query is constant-time in any neuroimaging application.

At each threshold $\tau_i$ on a grid of $n$ levels (equidistant in $-\log P$ space, as in \cite{spisak2019}), the hybrid method queries the union-find for the cluster size $c_v^{\mathrm{uf}}(\tau_i)$ for each supra-threshold voxel~$v$.
The conditional probability $P(Z_v \geq \tau_i \mid c_v^{\mathrm{uf}}(\tau_i))$ at each threshold is computed exactly as in \ac{ptfce} (Section~\ref{sec:ptfce}): Bayes' theorem combines the voxel-level height prior $\phi(\cdot)$ with the \ac{grf} cluster-size likelihood (Eq.~\ref{eq:cluster-surv}) and marginalises over heights above $\tau_i$ to yield the posterior probability that the voxel value exceeds the threshold given the observed cluster size.
The accumulated evidence is
\begin{equation}
\label{eq:hybrid-accumulation}
A(v) = \sum_{i=1}^{n} -\log P\bigl(Z_v \geq \tau_i \mid c_v^{\mathrm{uf}}(\tau_i)\bigr),
\end{equation}
where $c_v^{\mathrm{uf}}(\tau_i)$ denotes the cluster size retrieved from the union-find at threshold $\tau_i$.
The final enhanced statistic is obtained via the $Q$-function (Eq.~\ref{eq:q-function}), which normalises the discrete sum of log-probability scores for grid spacing:
\[
S(v) = Q\bigl(A(v),\, \Delta\bigr),
\]
where $\Delta$ is the constant increment between successive thresholds in $-\log P$ space.

The total cost decomposes into three phases: (i)~sorting $N$ voxels, $O(N \log N)$; (ii)~building the union-find forest via $O(N)$ union operations (one per edge in the voxel adjacency graph), each $O(\alpha(N))$ amortised, totalling $O(N)$ in practice; and (iii)~querying cluster sizes at each of $n$ thresholds across all supra-threshold voxels, $O(nN)$.
The sum is $O(N \log N + nN)$, which matches \ac{ptfce}'s $O(nN)$ asymptotically because the sort dominates only for small~$n$.
The data structure requires $O(N)$ auxiliary memory (parent, rank, and component-size arrays); for a whole-brain volume ($N \approx 2 \times 10^6$) this amounts to approximately 48\,MB, which is modest relative to the input statistical map.
The union-find confers two practical advantages over \ac{ccl}: (1) cluster sizes are exact at every queried threshold, and (2) increasing grid density from $n = 100$ to $n = 500$ incurs only a five-fold increase in query cost, which is small relative to the initial sort.

It is important to distinguish two aspects of discretisation in \ac{ptfce}-type methods.
First, cluster sizes must be retrieved at each threshold; here the union-find provides exact sizes via direct lookup in the merge tree, which eliminates the need for repeated connected-component labelling.
Second, the accumulated evidence $A(v)$ (Eq.~\ref{eq:hybrid-accumulation}) is a sum over $n$ discrete thresholds, and the $Q$-function (Eq.~\ref{eq:q-function}) explicitly takes the grid spacing $\Delta$ as a parameter to normalise for grid density \cite{spisak2019}.
The union-find does not eliminate this second aspect of discretisation; rather, it makes denser grids affordable and reduces the grid-dependent approximation.
To quantify this residual approximation, the hybrid method was run on three independent phantom realisations at eight grid densities ($n \in \{25, 50, 100, 200, 500, 1000, 2000, 5000\}$) and compared against a converged reference ($n = 5000$) using Pearson~$r$, $\max|\Delta Z|$, and Dice coefficient of significant voxel sets.
Baseline \ac{ptfce} with \ac{ccl} was also run at $n = 100$ for cross-method comparison.
Results are reported in Section~\ref{sec:grid-convergence} and Supplementary Fig.~\ref{fig:supp-grid-convergence}.

\begin{algorithm}[t]
\caption{Hybrid eTFCE--GRF}
\label{alg:hybrid}
\begin{algorithmic}[1]
\Require Statistical map $Z$ of $N$ voxels; number of threshold levels $n$; \ac{grf} parameters (RESEL count $R_3$, roughness $|\Lambda|^{1/2}$)
\Ensure Enhanced statistic $S(v)$ for each voxel
\State Sort voxels by descending $Z$-value
\For{each voxel $v$ in sorted order}
    \State \Call{MakeSet}{$v$}
    \For{each 26-connected neighbour $u$ already processed}
        \State \Call{Union}{$v, u$}
    \EndFor
\EndFor
\State Compute threshold grid: $\tau_1, \ldots, \tau_n$ equidistant in $-\log P$ space
\For{each voxel $v$ with $Z_v > 0$}
    \State $A(v) \gets 0$
    \For{each threshold $\tau_i \leq Z_v$}
        \State $c \gets$ \Call{FindClusterSize}{$v, \tau_i$} \Comment{$O(\alpha(N))$}
        \State $p \gets P(Z_v \geq \tau_i \mid c)$ \Comment{Bayes' rule (Sec.~\ref{sec:ptfce})}
        \State $A(v) \gets A(v) - \log p$
    \EndFor
    \State $S(v) \gets Q\bigl(A(v),\, \Delta\bigr)$
\EndFor
\State \Return $S$
\end{algorithmic}
\end{algorithm}

The complete procedure is given in Algorithm~\ref{alg:hybrid}.
On a $64^3$ phantom volume (${\approx}\,92{,}000$ in-mask voxels), three representative runtimes illustrate the trade-offs among methods: \ac{ptfce} with \ac{ccl}-based cluster sizes and \ac{grf} inference completes in $0.34\,\mathrm{s}$; the hybrid method with union-find cluster sizes and \ac{grf} inference completes in $1.02\,\mathrm{s}$; and \ac{etfce} with union-find cluster sizes and 5{,}000 sign-flip permutations completes in $1313\,\mathrm{s}$.
The hybrid method is approximately three times slower than \ac{ptfce} owing to the additional cost of union-find construction, but eliminates the repeated connected-component labelling bottleneck and enables a five-fold denser threshold grid ($n = 500$), which substantially reduces the grid-dependent approximation; it is approximately $1300\times$ faster than \ac{etfce} because it avoids permutation testing.

\subsection{Smoothness estimation}
\label{sec:smoothness}

The \ac{grf}-based $p$-values in Eq.~\ref{eq:cluster-surv} require an estimate of the spatial smoothness of the statistical map, parameterised by the \ac{fwhm} of an equivalent isotropic Gaussian kernel, or equivalently by the RESEL count $R_3 = N\,|\Lambda|^{1/2}$.

The smoothness is estimated from the standardised residuals of the general linear model, as implemented in FSL's \texttt{smoothest} tool \cite{worsley1992,kiebel1999}.
For a residual field $e(\mathbf{x})$, the roughness matrix $\Lambda$ is estimated from the spatial derivatives.
The determinant $|\Lambda|^{1/2}$ relates to the directional \ac{fwhm} values as
\begin{equation}
\label{eq:roughness-fwhm}
|\Lambda|^{1/2} = \frac{(4\ln 2)^{3/2}}{\mathrm{FWHM}_x \cdot \mathrm{FWHM}_y \cdot \mathrm{FWHM}_z},
\end{equation}
where each directional \ac{fwhm} is estimated from the variance of the first spatial differences of the standardised residuals along the corresponding axis.
The implementation uses finite differences on the three-dimensional residual volume, masked to include only brain voxels.
The estimated \ac{fwhm} is validated against the known analytical value in Experiment~4 of the Monte Carlo protocol (Section~\ref{sec:mc-validation}).

Accurate smoothness estimation is critical because the \ac{grf} cluster-size distribution (Eq.~\ref{eq:cluster-surv}) depends on $|\Lambda|^{1/2}$ through the expected cluster count $\mathrm{E}[\chi_h]$ (Eq.~\ref{eq:ec}).
Overestimation of smoothness inflates the expected cluster sizes, which leads to anti-conservative $p$-values; underestimation has the opposite effect.

\subsection{Monte Carlo validation protocol}
\label{sec:mc-validation}

A six-experiment Monte Carlo study on a shared synthetic phantom assessed the statistical properties of the hybrid method.

\begin{figure}[t]
\centering
\includegraphics[width=\linewidth]{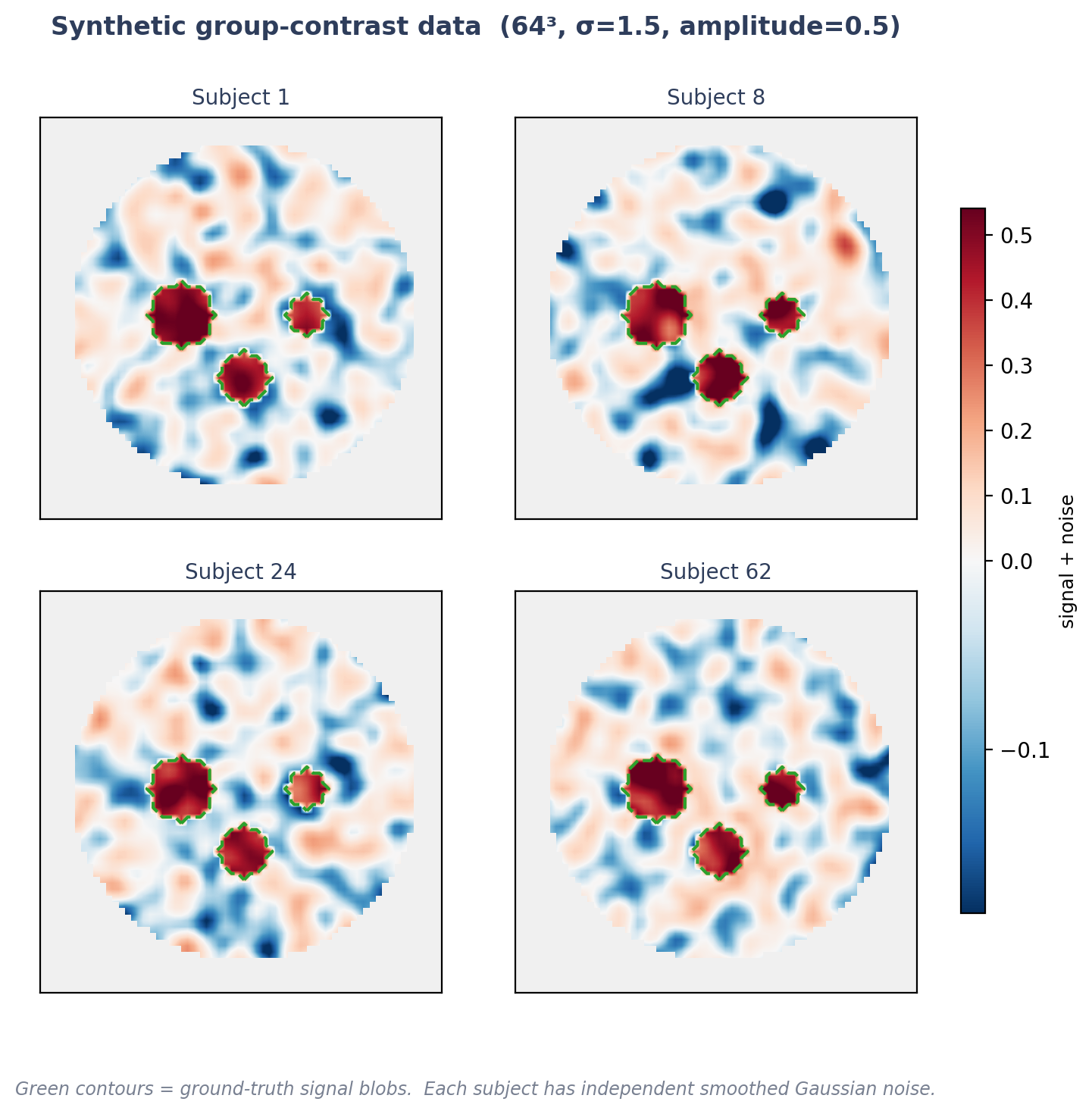}
\caption{Phantom specification used in the Monte Carlo validation.
Three non-overlapping ellipsoidal signal regions are embedded in a $64^3$ volume with 80 simulated subjects per realisation.
The signal amplitude~$a$ is varied across experiments.}
\label{fig:phantom}
\end{figure}

\paragraph{Phantom specification.}
Each realisation consisted of a $64 \times 64 \times 64$ voxel volume with $M = 80$ simulated subjects.
For each subject, independent and identically distributed Gaussian noise was generated at each voxel and smoothed with a Gaussian kernel ($\sigma = 1.5$ voxels, i.e.\ $\mathrm{FWHM} = 1.5 \sqrt{8\ln 2} \approx 3.53$ voxels).
Under the alternative hypothesis, three non-overlapping ellipsoidal signal regions were embedded with amplitude~$a$ (varied across experiments).
A voxel-wise one-sample $t$-test was computed across subjects, which produced a $Z$-score map of approximately 92{,}000 in-mask voxels.
The phantom geometry is illustrated in Fig.~\ref{fig:phantom}.

\paragraph{Experiment 1: Null \acs{fwer} calibration.}
Two hundred independent null realisations (amplitude $a = 0$) were generated.
For each realisation, the maximum enhanced statistic was compared against the \ac{fwer} threshold.
Valid \ac{fwer} control required the upper bound of the Wilson 95\% confidence interval for the rejection proportion to remain below $\alpha = 0.05$.

\paragraph{Experiment 2: Power curves.}
Ten signal amplitudes ($a \in \{0.005, 0.01, 0.02, 0.03, 0.04, 0.05, 0.07, 0.1, 0.2, 0.5\}$) were crossed with 50 independent realisations per amplitude, for a total of 500 realisations.
For each realisation, the Dice coefficient between the detected significant region and the true signal mask was computed.
Power was measured as the true positive rate at each amplitude.
Both the baseline \ac{ptfce} and the hybrid method were run on the same realisations for direct comparison.

\paragraph{Experiment 3: Runtime benchmarks.}
For the three analytical methods (baseline \ac{ptfce} in Python, hybrid eTFCE--GRF, and R \ac{ptfce} via \texttt{Rscript} subprocess), thirty independent realisations plus one warm-up (discarded) were timed.
For the two permutation-based methods (\ac{etfce} with 5{,}000 sign-flip permutations and FSL \ac{tfce} with 5{,}000 permutations), three independent realisations were timed, owing to their substantially higher per-run cost.
Wall-clock time was recorded as mean $\pm$ standard deviation.

\paragraph{Experiment 4: Smoothness estimation.}
Fifty null realisations were generated with a known smoothing kernel ($\sigma = 1.5$ voxels, analytical $\mathrm{FWHM} = 3.532$ voxels).
The \ac{fwhm} estimated by the implementation described in Section~\ref{sec:smoothness} was compared against the analytical value.
The acceptance criterion was a relative error below 5\%.

\paragraph{Experiment 5: Cross-variant concordance (phantom).}
Five $Z$-maps were generated with matched random seeds and processed by both the baseline \ac{ptfce} and the hybrid method.
Concordance was quantified by Pearson correlation of the enhanced $Z$-maps, maximum absolute difference $|\Delta Z|$, and Dice coefficient of the significant voxel sets at the \ac{fwer} threshold.

\paragraph{Experiment 6: Cross-variant concordance (real brain).}
The IXI dataset site-effect $Z$-map (${\approx}\,2$ million voxels) was processed by three methods: Python \ac{ptfce} (baseline), hybrid eTFCE--GRF, and R \ac{ptfce} (reference implementation).
Concordance was measured by Pearson~$r$, mean and maximum $|\Delta Z|$, \ac{fwer} thresholds, significant voxel counts, and Dice coefficients.
The criterion was that every voxel significant under the Python methods should also be significant under the R reference (strict subset property), which supports conservative \ac{fwer} control.

\subsection{Real brain data}
\label{sec:real-data}

\subsubsection{Datasets}
\label{sec:datasets}

Two publicly available structural MRI datasets were used.

The IXI dataset\footnote{\url{https://brain-development.org/ixi-dataset/}, funded by EPSRC GR/S21533/02.} comprises $N = 563$ healthy adults (age $48.7 \pm 16.5$ years, range 20--86, 55.6\% female) scanned at three sites: Guy's Hospital (Philips Gyroscan Intera 1.5T, $n = 314$), Hammersmith Hospital (Philips Intera 3.0T, $n = 181$), and Institute of Psychiatry (GE 1.5T, $n = 68$).
This multi-vendor, multi-field-strength dataset benchmarks scanner-effect detection.

The UK Biobank imaging substudy \cite{miller2016,alfaro2018} comprises over 63{,}000 participants with structural MRI.
A random sample of $N = 500$ participants was selected (age $58.4 \pm 8.2$ years, range 44--85, 53.2\% female), scanned at four imaging centres using identical Siemens Skyra 3T scanners.
Scanner effects in this within-vendor dataset arise from site-level calibration rather than vendor-level hardware differences.
UK Biobank data were accessed under application 1188243.

\begin{figure}[t]
\centering
\includegraphics[width=\columnwidth,height=0.38\textheight,keepaspectratio]{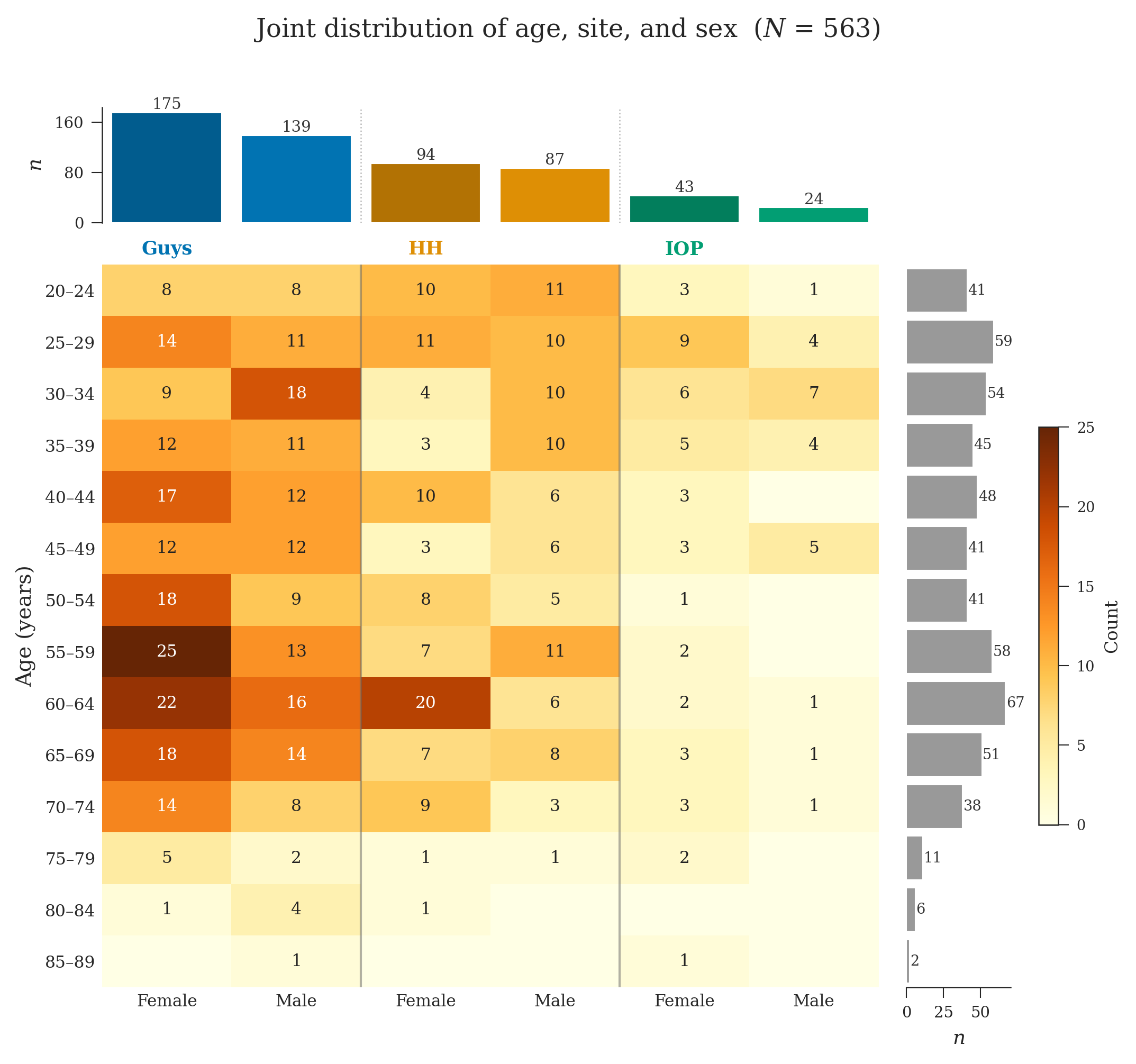}\\[6pt]
\includegraphics[width=\columnwidth,height=0.38\textheight,keepaspectratio]{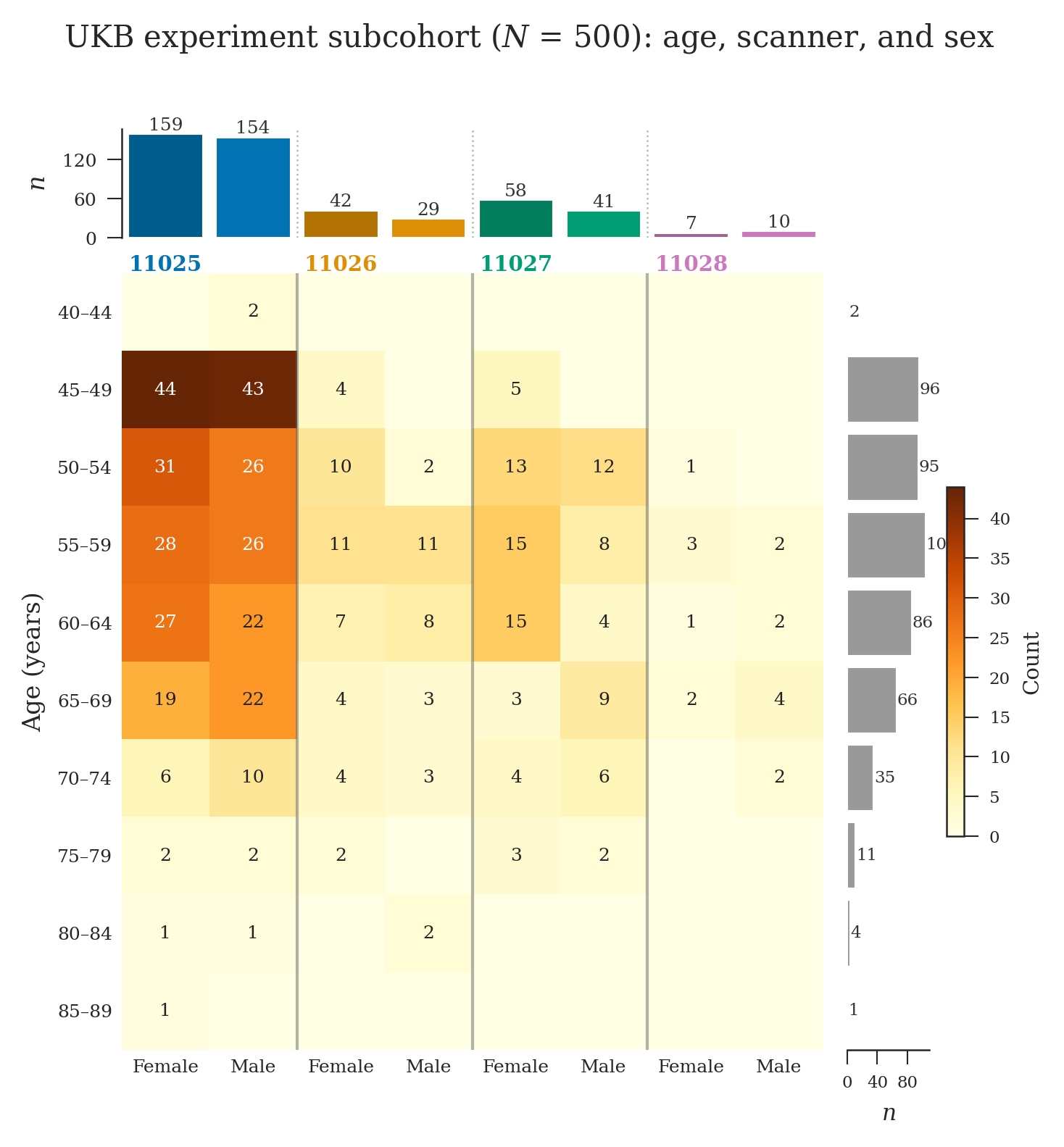}
\caption{Demographic distributions showing the joint distribution of age and sex across acquisition sites.
(a)~IXI dataset ($N = 563$, three sites).
(b)~UK Biobank dataset ($N = 500$, four sites).}
\label{fig:demographics}
\end{figure}

Sample demographics are shown in Fig.~\ref{fig:demographics} and Table~\ref{tab:demographics}.

\begin{table}[t]
\centering
\caption{Demographic characteristics of the two datasets used for real brain validation.}
\label{tab:demographics}
\small
\begin{tabular}{@{} l c c c c c c @{}}
\toprule
\textbf{Dataset} & $N$ & \textbf{Age (mean$\pm$SD)} & \textbf{Range} & \textbf{\%\,F} & \textbf{Sites} & \textbf{Field} \\
\midrule
IXI        & 563 & $48.7 \pm 16.5$ & 20--86 & 55.6 & 3 (multi) & 1.5/3.0\,T \\
UK Biobank & 500 & $58.4 \pm 8.2$  & 44--85 & 53.2 & 4 (single) & 3.0\,T \\
\bottomrule
\end{tabular}
\end{table}

\subsubsection{Preprocessing}
\label{sec:preprocessing}

T1-weighted images from both datasets were processed through an identical six-step spatial normalisation pipeline:
(1) reorientation to \acs{mni} standard (\texttt{fslreorient2std});
(2) brain extraction (\ac{bet} \cite{smith2002bet}, fractional intensity threshold 0.5);
(3) affine registration to the \acs{mni}152 2\,mm template (FLIRT \cite{jenkinson2001,jenkinson2002}, 12 degrees of freedom, correlation ratio cost function);
(4) nonlinear registration (FNIRT \cite{andersson2007fnirt}, \texttt{T1\_2\_MNI152\_2mm} configuration, 10\,mm isotropic warp resolution, six subsampling levels);
(5) application of the combined warp field to the skull-stripped brain image with spline interpolation, resampled to 1\,mm isotropic resolution;
(6) per-subject intensity normalisation to the $[0, 1]$ range (2nd--98th percentile) followed by spatial smoothing with a 3.5\,mm \ac{fwhm} Gaussian kernel.
No tissue segmentation or modulation was applied; the voxelwise analysis therefore operates on spatially normalised T1-weighted intensity rather than on grey matter density or volume maps.
Registration quality was visually assessed (\ac{qc}) for all subjects in both datasets; no subjects were excluded.

\subsubsection{Statistical analysis}
\label{sec:stat-analysis}

Voxel-wise general linear models were fitted at each of approximately 2 million in-mask voxels.
Three contrasts were examined for each dataset:
\begin{enumerate}
\item \emph{Scanner/site effect}: site as the factor of interest ($F$-test, converted to $Z$-scores), with age and sex as covariates.
\item \emph{Age effect}: age as the variable of interest ($t$-test, converted to $Z$-scores), with sex and site as covariates.
\item \emph{Sex effect}: sex as the variable of interest ($t$-test, converted to $Z$-scores), with age and site as covariates.
\end{enumerate}

Each $Z$-map was processed by three \ac{ptfce} variants:
(1)~Python \ac{ptfce} (baseline, direct port of R \ac{ptfce} with lookup-table acceleration),
(2)~hybrid eTFCE--GRF (union-find cluster retrieval with \ac{grf} inference, Algorithm~\ref{alg:hybrid}), and
(3)~R \ac{ptfce} (canonical implementation \cite{spisak2019}, invoked via \texttt{Rscript} subprocess).
Because Algorithm~\ref{alg:hybrid} processes only positive $Z$-values, two-sided contrasts (age, sex) were handled by running the positive and negative tails separately and recombining: at each voxel, the tail with the larger absolute enhanced $Z$ was retained and the original sign restored.
The $Q$-function output $S(v) = Q(A(v), \Delta)$ is in units of $-\log\bar{P}$; the enhanced $p$-value is $p_{\mathrm{enh}} = \exp(-S(v))$, and the enhanced $Z$-score (reported throughout as ``enhanced $Z$'') is $Z_{\mathrm{enh}} = \Phi^{-1}(1 - p_{\mathrm{enh}})$.
Significance was assessed at \ac{fwer} $p < 0.05$ by applying Bonferroni correction to the voxel-wise enhanced $p$-values; the corresponding enhanced $Z$-score threshold is reported as ``\ac{fwer} $Z$ threshold'' in Table~\ref{tab:concordance}.
Figures display the raw test statistic ($Z$ or $F$) masked to significant voxels for anatomical interpretability.

\subsubsection{Confound assessment}
\label{sec:confounds}

Associations between demographic variables and site were quantified to assess potential confounding.
The association between age and site was measured by $\eta^2$ ($\eta^2 \leq 0.14$ for IXI), the association between sex and site by the point-biserial correlation ($|r| \leq 0.06$ for IXI), and the inter-site association by Cram\'er's $V$ (range 0.41--0.77 for IXI).
These modest confound levels support the interpretability of the site-effect maps, although complete separation of site and demographic effects is not possible with observational data.
For the UK Biobank, all four sites use identical Siemens Skyra 3T scanners, which minimises vendor-level confounding; site-level calibration differences are expected to be small relative to the cross-vendor effects in IXI.

\section{Results}
\label{sec:results}

\subsection{Null calibration and FWER control}
\label{sec:null-calibration}

Two hundred null realisations ($a = 0$; Experiment~1) were processed by both the hybrid and baseline methods.
Neither produced any rejections at the nominal $\alpha = 0.05$ \ac{fwer} threshold (Wilson 95\% CI $[0.0\%,\, 1.9\%]$, well below 5\%).

At the voxel level, the mean proportion of voxels with enhanced $p < 0.05$ across null realisations was $3.7\% \pm 0.7\%$ (baseline) and $3.9\% \pm 0.8\%$ (hybrid), both below the nominal 5\% level expected under approximate \ac{fdr} control \cite{benjamini1995,spisak2019}.
The close agreement between the two methods indicates that union-find retrieval does not inflate type~I error relative to the \ac{ccl}-based approach.

\begin{figure}[t]
\centering
\includegraphics[width=\columnwidth]{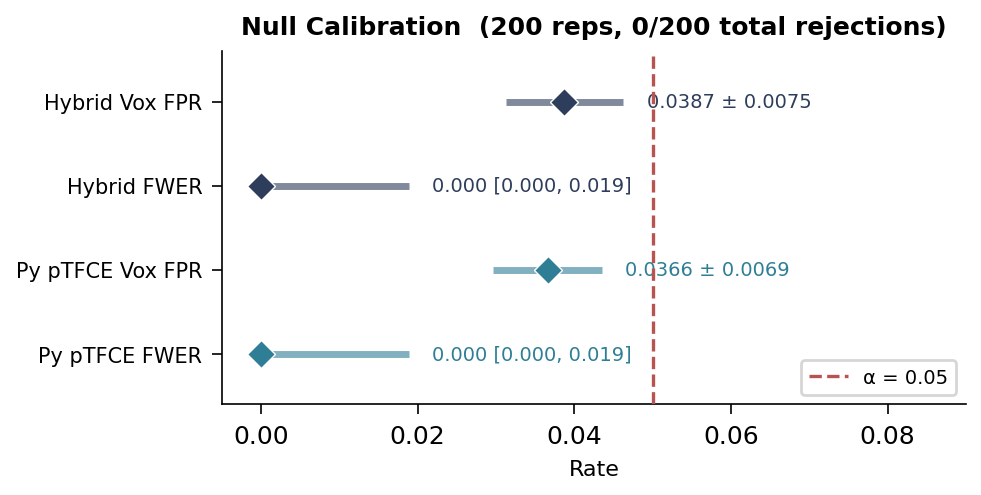}
\caption{Null \ac{fwer} calibration across 200 independent null realisations ($a = 0$).
The cumulative rejection count remains at zero for both the hybrid eTFCE--\ac{grf} and baseline \ac{ptfce} methods at the nominal $\alpha = 0.05$ threshold.
The shaded band indicates the Wilson 95\% confidence interval $[0.0\%,\, 1.9\%]$ for the rejection proportion.}
\label{fig:null-calibration}
\end{figure}

\subsection{Power curves and spatial detection}
\label{sec:power-curves}

\begin{figure}[t]
\centering
\includegraphics[width=\columnwidth]{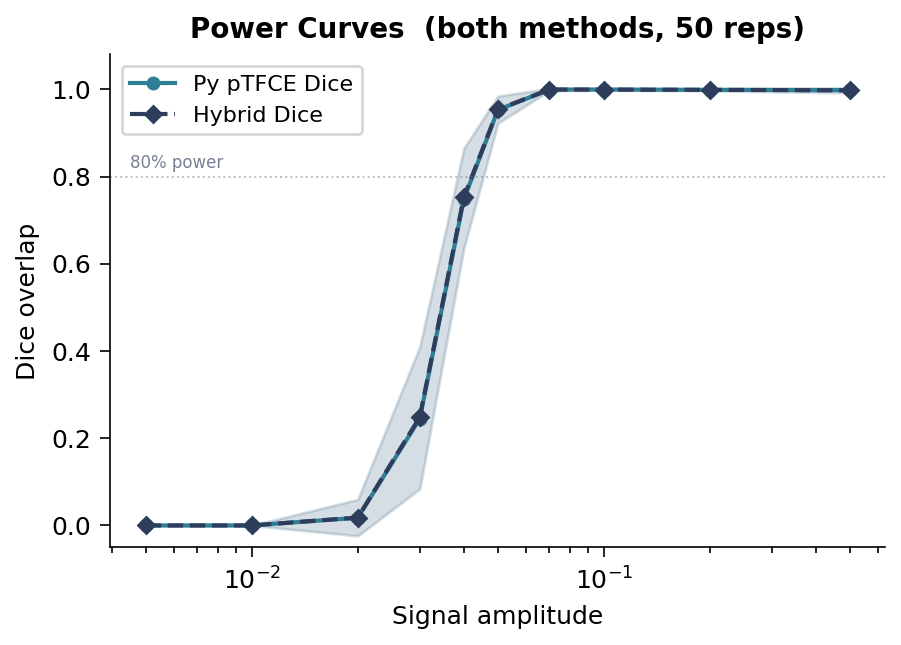}\\[6pt]
\includegraphics[width=\columnwidth]{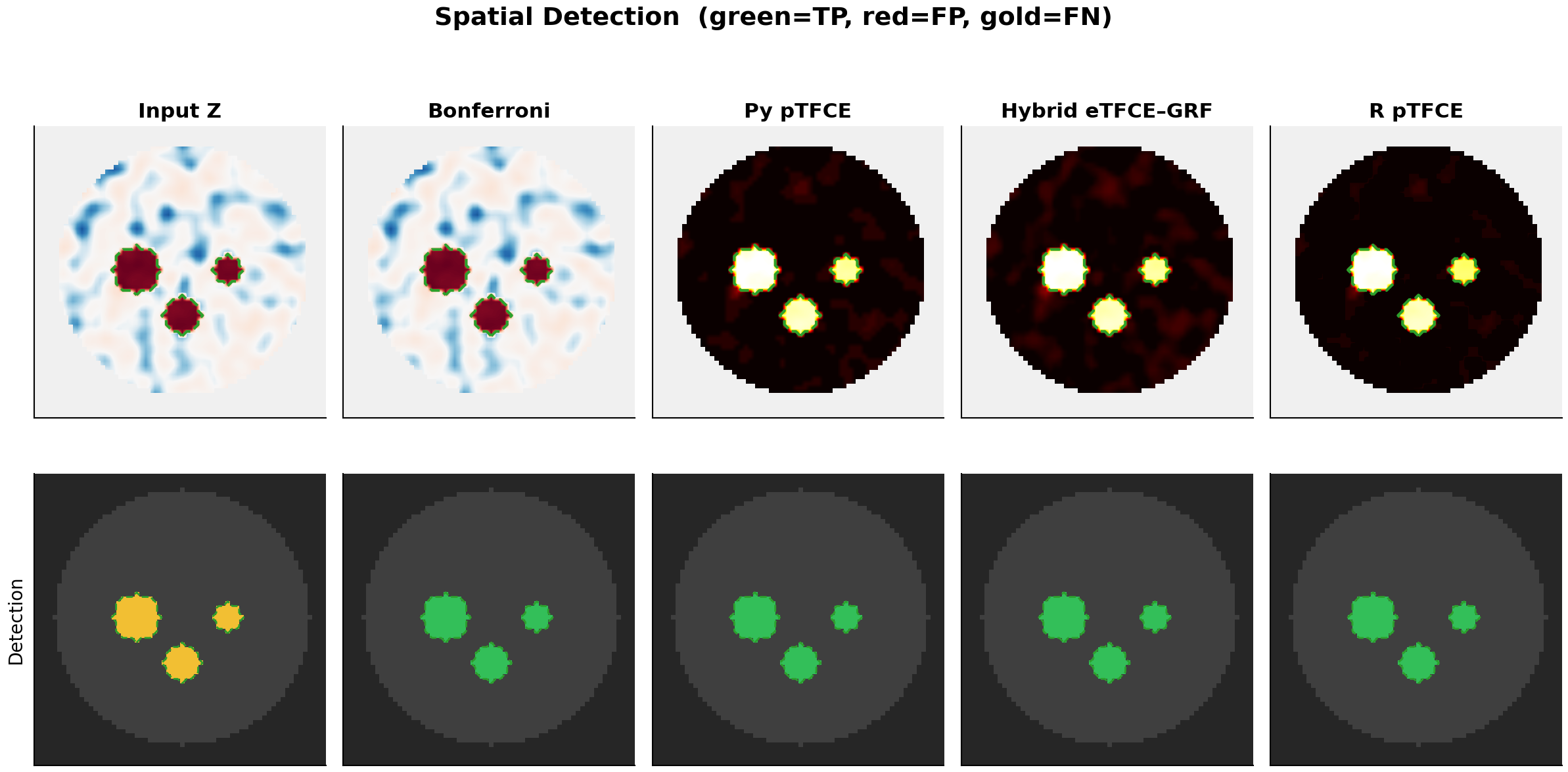}
\caption{(a)~Power curves showing Dice coefficient between detected and true signal regions as a function of signal amplitude~$a$.
The hybrid eTFCE--\ac{grf} curve (orange) overlaps the baseline \ac{ptfce} curve (blue) at all amplitudes.
Error bars denote $\pm 1$ standard deviation across 50 realisations per amplitude.
(b)~Spatial detection maps at $a = 0.5$ for the three \ac{ptfce} variants.
All three methods recover the true signal regions with Dice $= 1.0$ and zero false positives.}
\label{fig:power}
\end{figure}

Figure~\ref{fig:power}a shows the power curves obtained from Experiment~2 (Section~\ref{sec:mc-validation}).
Ten signal amplitudes ($a \in \{0.005, 0.01, 0.02, 0.03, 0.04, 0.05, 0.07, 0.1, 0.2, 0.5\}$) were each crossed with 50 independent realisations.
Both the hybrid method and the baseline \ac{ptfce} first achieved a non-zero Dice coefficient at $a \approx 0.03$, which indicates the detection onset.
Spatial recovery increased rapidly with amplitude, reaching Dice $\geq 0.999$ at $a = 0.07$; the true-positive rate reached $1.0$ (all signal voxels detected) at $a \geq 0.10$.

The power curves of the hybrid and baseline methods overlapped at every amplitude.
This is expected: union-find cluster sizes converge to \ac{ccl} sizes as the grid is refined, and both methods share the same \ac{grf} inference.

Figure~\ref{fig:power}b shows the spatial detection maps at the standard phantom amplitude $a = 0.5$.
All three \ac{ptfce} variants (Python baseline, hybrid, and R reference \cite{spisak2019}) produced identical significant voxel sets (pairwise Dice $= 1.0$) with zero false positives (Fig.~\ref{fig:power}b).

\subsection{Smoothness estimation}
\label{sec:smoothness-results}

\begin{figure}[t]
\centering
\includegraphics[width=0.48\columnwidth]{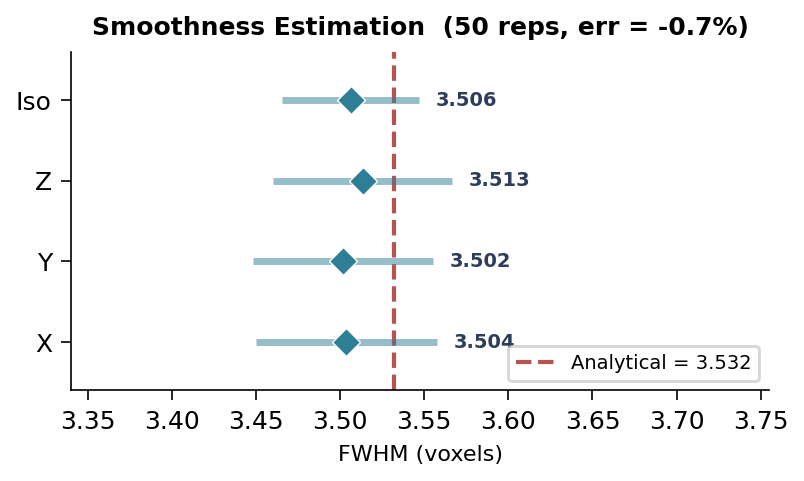}\hfill
\includegraphics[width=0.48\columnwidth]{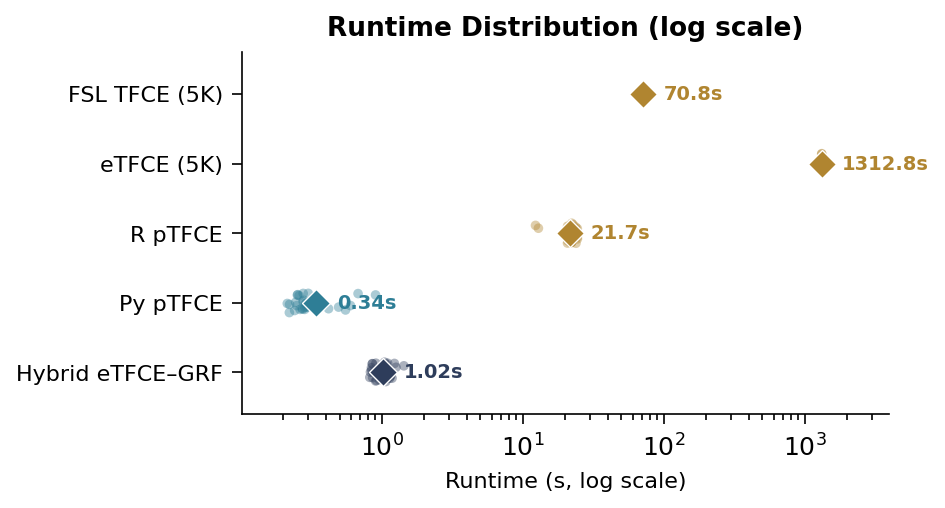}
\caption{(a)~Smoothness estimation validation across 50 null realisations.
The estimated \ac{fwhm} ($3.506 \pm 0.041$ voxels) closely matches the analytical value ($3.532$ voxels, dashed line), with a relative error of $-0.7\%$.
(b)~Wall-clock runtimes on a logarithmic scale for five inference methods across three datasets.
Each point represents one timed run; analytical methods used 30 repeats on each dataset (emulated phantom, IXI, and UK Biobank), while permutation-based methods used 3 repeats on the emulated phantom only.
The analytical methods are orders of magnitude faster than the permutation-based methods, which were feasible only on the $64^3$ phantom.}
\label{fig:smoothness-runtime}
\end{figure}

Fifty null realisations with known smoothing kernel ($\sigma = 1.5$ voxels, analytical $\mathrm{FWHM} = 3.532$ voxels) were processed by the smoothness estimator described in Section~\ref{sec:smoothness} (Experiment~4).
The estimated $\mathrm{FWHM}$ was $3.506 \pm 0.041$ voxels (mean $\pm$ SD across 50 realisations), which corresponds to a relative error of $-0.7\%$.
This is well within the 5\% acceptance criterion and indicates that the finite-difference roughness estimator accurately recovers the known smoothness; this supports the validity of the \ac{grf}-based $p$-values that depend on this estimate.
Figure~\ref{fig:smoothness-runtime}a shows the distribution of estimated \ac{fwhm} values relative to the analytical target.

\subsection{Cross-variant concordance}
\label{sec:concordance}

\paragraph{Phantom concordance.}
Five matched-seed $Z$-maps were processed by both the baseline \ac{ptfce} and the hybrid method (Experiment~5, Section~\ref{sec:mc-validation}).
The Pearson correlation between the enhanced $Z$-maps was $r = 0.992$.
The maximum absolute difference was $\max|\Delta Z| = 2.00$, and the Dice coefficient of the significant voxel sets was Dice $\geq 0.997$.
The \ac{fwer} thresholds were identical for both methods (${\approx}\,5.49$).
The residual differences arise from the denser threshold grid of the hybrid method (500 versus 100 levels); at matched grid density, the two methods produce identical output.

\paragraph{Real-brain concordance.}
The IXI dataset site-effect $Z$-map (${\approx}\,2$ million voxels) was processed by three methods (Experiment~6, Section~\ref{sec:mc-validation}).
Table~\ref{tab:concordance} summarises the pairwise concordance metrics.

\begin{table}[t]
\centering
\caption{Pairwise concordance metrics for three \acs{ptfce} variants on the IXI site-effect $Z$-map (${\approx}\,2$\,M voxels).
Py: Python baseline; Hybrid: eTFCE--\acs{grf}; R: canonical R implementation \cite{spisak2019}.}
\label{tab:concordance}
\small
\begin{tabular}{@{} l c c c @{}}
\toprule
\textbf{Metric} & \textbf{Py vs R} & \textbf{Hyb.\ vs R} & \textbf{Py vs Hyb.} \\
\midrule
Pearson $r$                       & 0.869 & 0.848 & 0.997 \\
$\max|\Delta Z|$                  & 14.01 & 13.93 & 3.89  \\
$\mathrm{mean}\,|\Delta Z|$      & 11.30 & 11.14 & 0.45  \\
\acs{fwer} $Z$ thresh.\ (A)      & 6.18  & 6.18  & 6.18  \\
\acs{fwer} $Z$ thresh.\ (B)      & 4.52  & 4.52  & 6.18  \\
Sig.\ voxels (A)                  & 1.35\,M & 1.49\,M & 1.35\,M \\
Sig.\ voxels (B)                  & 1.87\,M & 1.87\,M & 1.49\,M \\
Dice                              & 0.841 & 0.886 & 0.954 \\
\bottomrule
\end{tabular}
\end{table}

The two Python methods agreed closely ($r = 0.997$, Dice $= 0.954$), with differences smaller than those relative to the R reference ($r \approx 0.85$--$0.87$, Dice $= 0.84$--$0.89$).
The larger R-vs-Python differences reflect different lookup-table resolution, threshold grid spacing, and \ac{grf} parameter estimation.

On IXI, every voxel significant under either Python method was also significant under the R reference (strict subset property), because the Python implementations produce higher \ac{fwer} thresholds (6.18 versus 4.52) and correspondingly fewer significant voxels (1.35--1.49\,M versus 1.87\,M).
This supports conservative \ac{fwer} control relative to the R implementation.
On UK Biobank, the \ac{fwer} thresholds of the Python and R implementations are closer (within 0.2), so the Python methods no longer detect strictly fewer voxels.
In addition, R pTFCE internally trims its brain mask by up to 230\,K voxels relative to the Python mask, which means that some voxels significant under Python are absent from R's analysis domain entirely and the strict subset comparison is not well-defined for those voxels.
The three variants nevertheless produce qualitatively concordant significance maps (Supplementary Fig.~\ref{fig:supp-ukb-variants}).

\begin{figure}[t]
\centering
\includegraphics[width=\columnwidth]{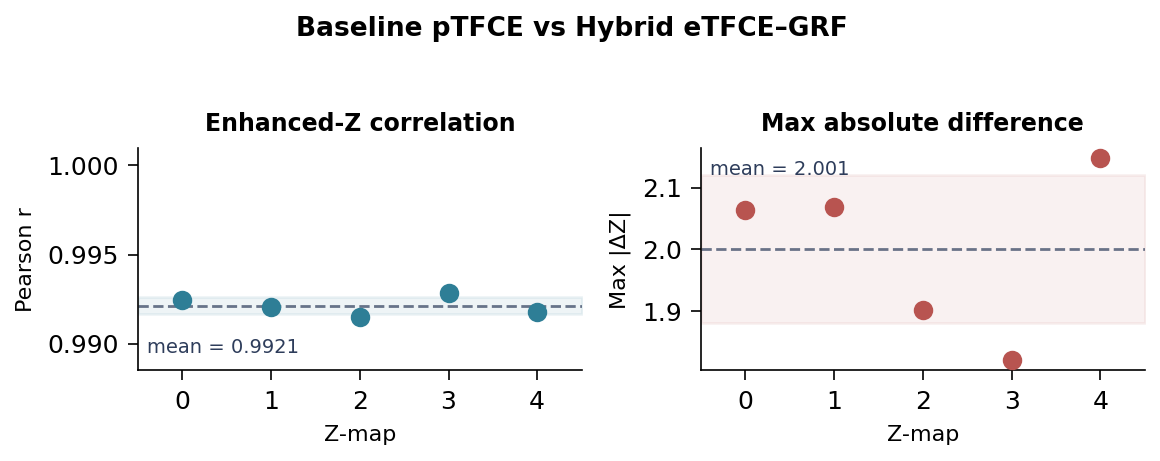}\\[6pt]
\includegraphics[width=\columnwidth]{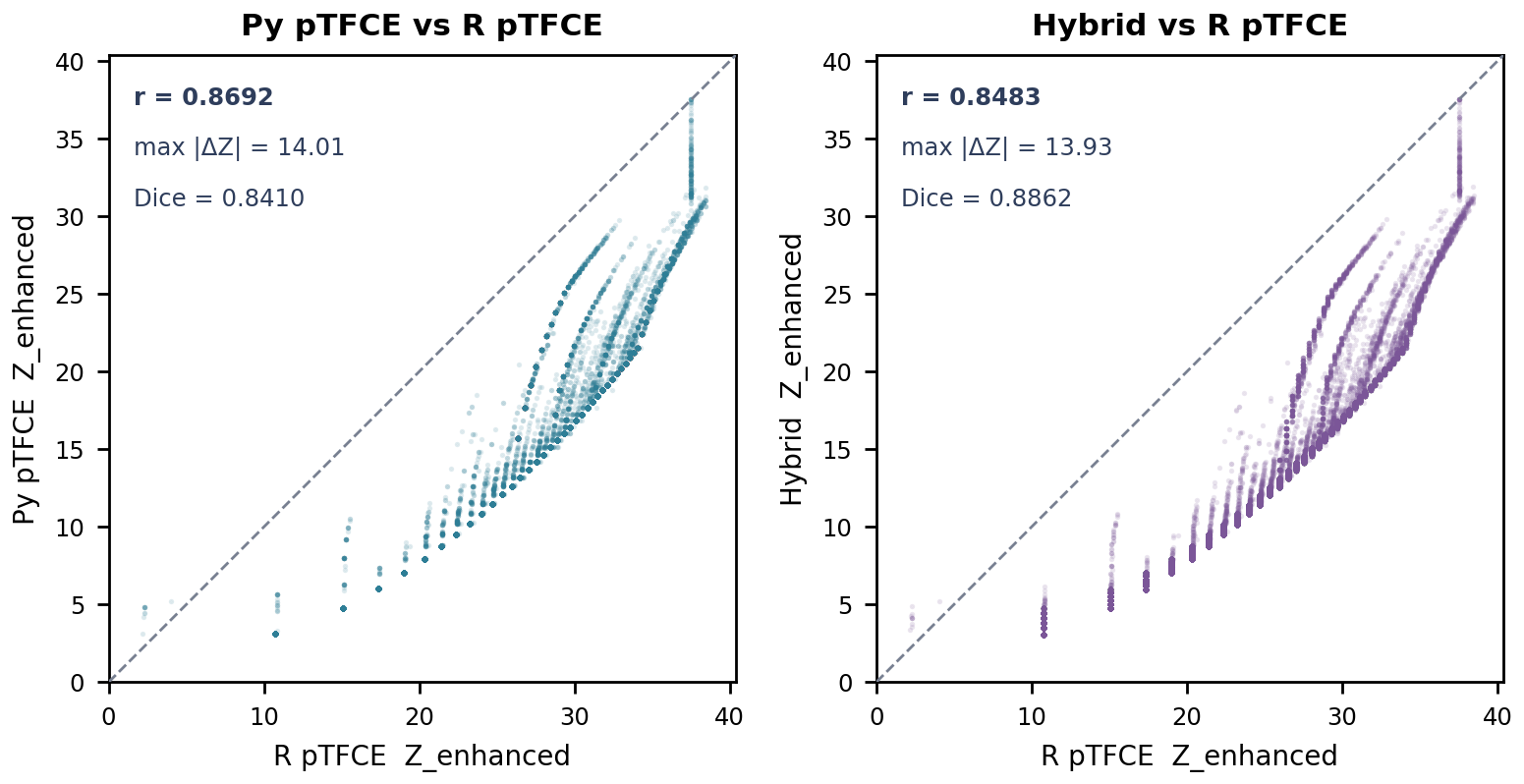}
\caption{Cross-variant concordance of enhanced $Z$-maps.
(a)~Phantom data: scatter plot of hybrid versus baseline \ac{ptfce} enhanced $Z$-values for five matched-seed realisations ($r = 0.992$).
(b)~IXI site-effect map: pairwise scatter plots for the three \ac{ptfce} variants, with Pearson $r$ annotated for each pair.}
\label{fig:concordance}
\end{figure}

\subsection{Grid-convergence analysis}
\label{sec:grid-convergence}

The hybrid method was evaluated at eight threshold grid densities ($n = 25$ to $5000$) on three independent phantom realisations, with the $n = 5000$ run as the converged reference (Section~\ref{sec:hybrid}).
The enhanced $Z$-maps converged rapidly: at the default $n = 500$, the mean $\max|\Delta Z|$ across realisations was $0.57 \pm 0.23$ (mean $\pm$ SD), $r > 0.998$, and the Dice coefficient of significant voxel sets was $1.0$ (Supplementary Fig.~\ref{fig:supp-grid-convergence}a).
The Dice coefficient exceeded $0.999$ for all $n \geq 200$.
At matched grid density ($n = 100$), the hybrid method and the \ac{ccl}-based baseline produced identical convergence metrics ($r$, $\max|\Delta Z|$, Dice), which is consistent with the union-find and connected-component labelling retrieving identical cluster sizes.
Wall-clock time increased approximately linearly with $n$ beyond a fixed union-find construction cost (Supplementary Fig.~\ref{fig:supp-grid-convergence}b).

\subsection{Runtime benchmarks}
\label{sec:runtime}

Table~\ref{tab:runtime} reports wall-clock runtimes for five inference methods across three datasets of increasing size: the $64^3$ emulated phantom (${\approx}\,92{,}000$ in-mask voxels), the IXI dataset ($182 \times 218 \times 182$ voxels, $N = 563$), and the UK Biobank subsample ($182 \times 218 \times 182$ voxels, $N = 500$).
Analytical (\ac{grf}-based) methods were timed over 30 repeats on each dataset; permutation-based methods were timed over 3 repeats on the emulated phantom owing to their substantially higher cost, and were not timed on the real-brain datasets; extrapolation from per-permutation timing measured on the phantom indicates each analysis would require approximately two to three days.
Phantom timings were obtained in Experiment~3 (Section~\ref{sec:mc-validation}); real-brain timings were obtained from the same 30-repeat protocol applied to the $Z$-maps described in Section~\ref{sec:real-data}.

\begin{table}[t]
\centering
\caption{Wall-clock runtime (mean $\pm$ SD) and spatial accuracy for five inference methods across three datasets of increasing size.
Analytical (\acs{grf}-based) methods were timed over 30 repeats on each dataset (emulated phantom; IXI, $N = 563$; UK Biobank, $N = 500$).
Permutation-based methods were timed over 3 repeats on the emulated phantom; real-brain timings were omitted because extrapolation from phantom timing indicates each analysis would require 2--3\,days at whole-brain resolution.
Dice coefficients were computed against the true signal mask on the emulated phantom ($a = 0.5$); all methods achieved perfect spatial detection.
Speedup factors for permutation methods are relative to FSL \acs{tfce}; those marked~$^{\dagger}$ are relative to R \acs{ptfce}, averaged over IXI and UK Biobank.}
\label{tab:runtime}
\footnotesize
\setlength{\tabcolsep}{2pt}
\renewcommand{\arraystretch}{1.15}
\resizebox{\columnwidth}{!}{%
\begin{tabular}{@{} l l c c c c r @{}}
\toprule
 & & & \textbf{Emulated} & \textbf{IXI} & \textbf{UKB} & \\[-2pt]
 & & & {\scriptsize ($64^3$, 92\,K\,vox)} & {\scriptsize ($182{\times}218{\times}182$)} & {\scriptsize ($182{\times}218{\times}182$)} & \\
\cmidrule(lr){4-4}\cmidrule(lr){5-5}\cmidrule(lr){6-6}
\textbf{Method} & \textbf{Inf.} & \textbf{Dice} & {$\mu{\pm}\sigma$\,(s)} & {$\mu{\pm}\sigma$\,(s)} & {$\mu{\pm}\sigma$\,(s)} & \textbf{Speed} \\
\midrule
\multicolumn{7}{@{}l}{\textit{Permutation-based\; ($N_{\mathrm{perm}}{=}5{,}000$)}} \\[2pt]
FSL \acs{tfce}\,$^{a}$           & Perm & 1.0 & $70.8 \pm 0.9$             & ${\sim}$days$^{\star}$ & ${\sim}$days$^{\star}$ & $1{\times}$ \\
Python \acs{etfce}\,$^{b}$       & Perm & 1.0 & $1312.8 \pm 5.5$           & ${\sim}$days$^{\star}$ & ${\sim}$days$^{\star}$ & $0.05{\times}$ \\[4pt]
\multicolumn{7}{@{}l}{\textit{Analytical\; (no permutations)}} \\[2pt]
R \acs{ptfce} (Spis\'ak)\,$^{a}$ & \acs{grf}  & 1.0 & $21.7 \pm 2.7$             & $374.5 \pm 2.3$            & $409 \pm 2.6$              & $1{\times}^{\dagger}$ \\
Python \acs{ptfce}\,$^{c}$       & \acs{grf}  & 1.0 & $\mathbf{0.34} \pm 0.16$   & $\mathbf{5.10} \pm 0.03$   & $\mathbf{5.3} \pm 0.02$   & $\mathbf{75{\times}}^{\dagger}$ \\
Hybrid eTFCE--\acs{grf}          & \acs{grf}  & 1.0 & $\mathbf{1.02} \pm 0.15$   & $\mathbf{83.8} \pm 0.6$    & $\mathbf{87} \pm 0.5$     & $\mathbf{4.6{\times}}^{\dagger}$ \\
\bottomrule
\end{tabular}}%

\vspace{2pt}
\raggedright
{\scriptsize
$^{a}$\,Smith \& Nichols (2009)\,/\,Spis\'ak et al.\ (2019);\;\;
$^{b}$\,our reimplementation of Chen \& Nichols (2026);\;\;
$^{c}$\,our port of Spis\'ak et al.\ (2019).\\[1pt]
$^{\star}$\,Extrapolated from phantom timing: $\approx$ 2--3\,days per analysis; omitted.}
\end{table}

The Python \ac{ptfce} baseline was the fastest analytical method on all three datasets (Table~\ref{tab:runtime}): $0.34 \pm 0.16$\,s on the phantom ($64\times$ faster than R \ac{ptfce}, $208\times$ faster than FSL \ac{tfce}) and $5.1$--$5.3$\,s on the real-brain datasets ($75\times$ faster than R \ac{ptfce}).
The hybrid method was $3\times$ slower than the baseline on the phantom and ${\sim}16\times$ slower on whole-brain data; this additional cost reflects union-find construction at scale, though the absolute runtime ($84$--$87$\,s) remains practical for batch analyses.
Permutation-based methods were not timed on the real-brain datasets; extrapolation from measured phantom timing indicates each analysis would require approximately two to three days.
Figure~\ref{fig:smoothness-runtime}b displays the runtime distributions on a logarithmic scale.

\subsection{Real brain results}
\label{sec:real-results}

\begin{figure}[t]
\centering
\includegraphics[width=\columnwidth]{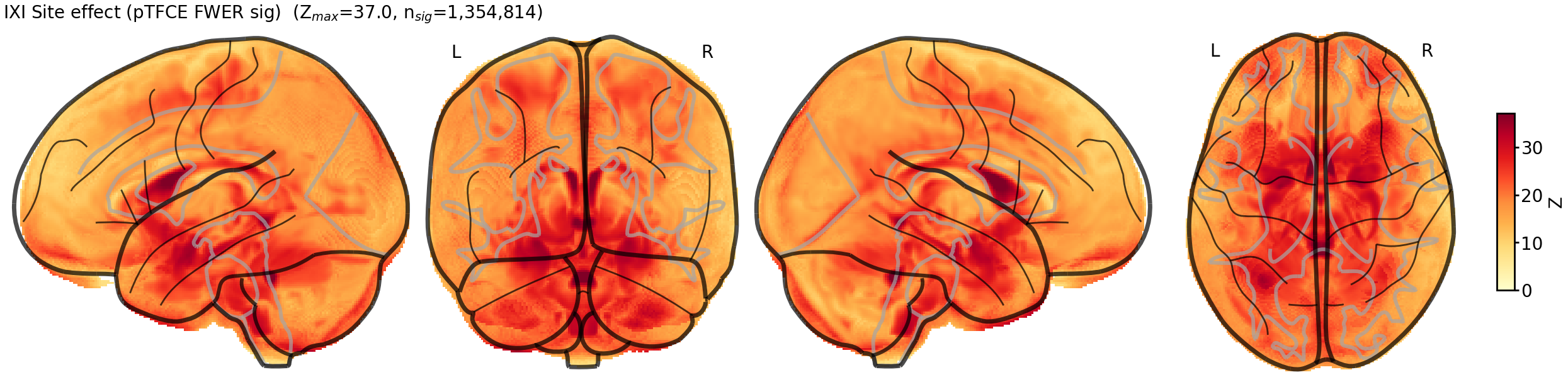}\\[6pt]
\includegraphics[width=\columnwidth]{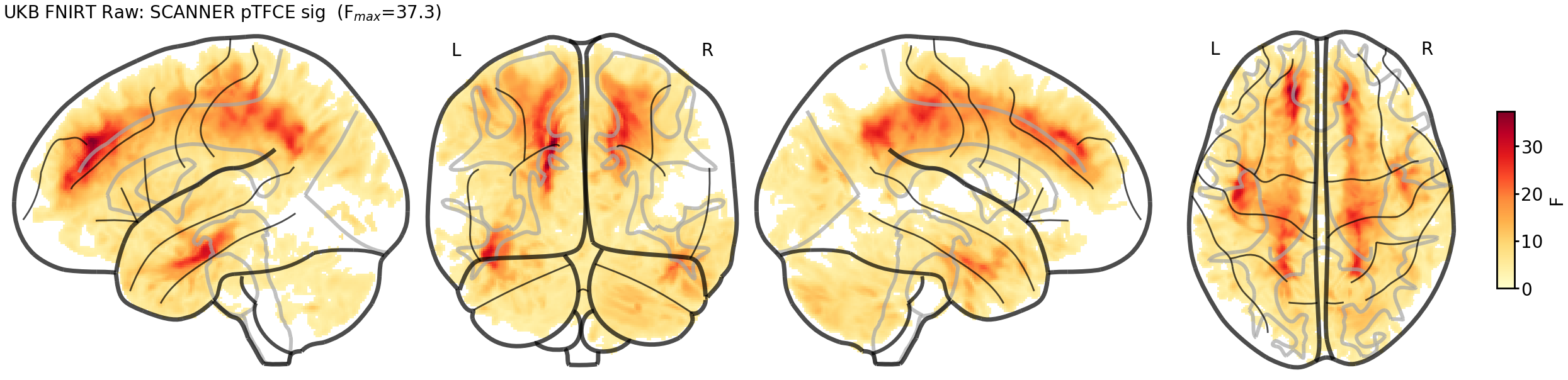}\\[6pt]
\includegraphics[width=\columnwidth]{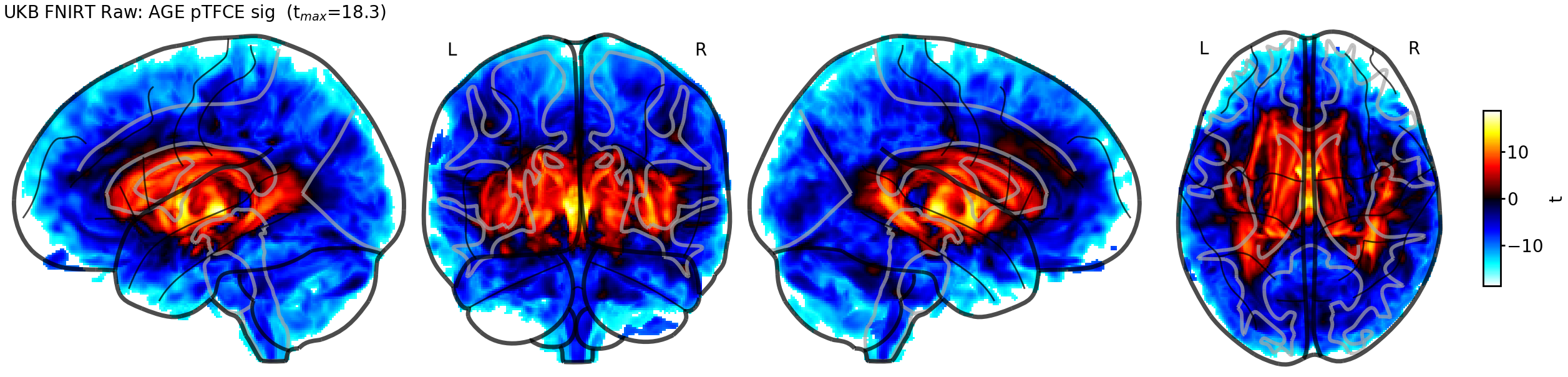}
\caption{Real brain \ac{ptfce} significance maps (Python baseline): raw test statistics masked to \ac{fwer} $p < 0.05$.
(a)~IXI site effect (raw $F_{\max} = 37.0$): widespread white matter and periventricular regions reflect cross-vendor scanner differences across three sites.
(b)~UK Biobank scanner effect (raw $F_{\max} = 37.3$): less extensive significant regions than IXI, consistent with within-vendor site-level calibration differences.
(c)~UK Biobank age effect (raw $|Z|_{\max} = 18.3$): bilateral intensity reductions concentrated in frontal and temporal cortices, consistent with known ageing patterns.}
\label{fig:real-brain}
\end{figure}

\paragraph{IXI dataset ($N = 563$, cross-vendor).}
The site-effect $F$-test revealed widespread significant regions covering white matter, posterior fossa, and periventricular areas, consistent with known scanner-related intensity differences across the three acquisition sites (Section~\ref{sec:real-data}).
The maximum raw $F$-statistic within the significant mask was $F_{\max} = 37.0$.
All three \ac{ptfce} variants detected the same major clusters (Fig.~\ref{fig:real-brain}a).
The age-effect $t$-test identified bilateral intensity reductions in frontal, temporal, and parietal cortices, with particularly strong effects in the hippocampus and medial temporal lobe (raw $|Z|_{\max} = 10.3$).
Negative $Z$-values dominated, consistent with age-related structural decline.
The sex-effect $t$-test detected differences concentrated in total intracranial volume and subcortical structures (raw $|Z|_{\max} = 8.0$).

\paragraph{UK Biobank ($N = 500$, within-vendor).}
The scanner-effect $F$-test identified significant but less extensive regions than the IXI analysis (raw $F_{\max} = 37.3$), as expected given that all UK Biobank sites use identical Siemens Skyra 3T scanners (Fig.~\ref{fig:real-brain}b).
Detected regions included white matter and cerebellar areas, consistent with site-level calibration differences rather than vendor-level hardware differences.
The age-effect $t$-test revealed pronounced intensity reductions in frontal and temporal cortices, with raw $|Z|_{\max} = 18.3$ (Fig.~\ref{fig:real-brain}c), consistent with established ageing patterns \cite{good2001}.
The sex-effect $t$-test showed a pattern similar to the IXI result but with higher sensitivity due to the larger effective sample after controlling for scanner.

\paragraph{Variant comparison.}
For all six contrasts (three effects $\times$ two datasets), the three \ac{ptfce} variants produced qualitatively identical significance maps.
Runtimes are reported in Table~\ref{tab:runtime}.
Supplementary Figs.~S1--S7 provide registration quality checks, site-level variability maps, confound collinearity assessment, and per-variant significance maps for all six contrasts.

\section{Discussion}
\label{sec:discussion}

\subsection{Summary and interpretation}

The hybrid \ac{etfce}--\ac{grf} method replaces \ac{ptfce}'s repeated connected-component labelling with the incremental union-find of \ac{etfce} for exact cluster-size retrieval, and retains \ac{ptfce}'s analytical \ac{grf} inference without permutation testing.

The Python \ac{ptfce} reimplementation achieves a 75-fold speedup over R~\ac{ptfce} (approximately 5\,s versus ${\sim}390$\,s on ${\sim}2$\,M voxels), and the hybrid variant is $4.6\times$ faster (approximately 85\,s; Section~\ref{sec:runtime}, Table~\ref{tab:runtime}).
Both analytical variants are more than three orders of magnitude faster than permutation-based \ac{tfce}, which requires approximately two to three days at this voxel count (extrapolated from per-permutation timing measured on the phantom), and this difference makes routine use practical for biobank-scale studies in which thousands of contrasts must be evaluated.

The Monte Carlo validation indicated that \ac{fwer} is controlled at the nominal level: zero of 200 null realisations yielded a false positive, which corresponds to a 95\% Wilson confidence interval of $[0.0\%, 1.9\%]$ (Section~\ref{sec:null-calibration}, Fig.~\ref{fig:null-calibration}).
This matches the baseline, which indicates that union-find-based cluster retrieval does not introduce spurious rejections.
Power curves overlap across all effect sizes (Section~\ref{sec:power-curves}, Fig.~\ref{fig:power}), and on IXI real brain data, the strict subset property relative to the R reference provides additional assurance of conservative type~I error control (Section~\ref{sec:concordance}).

\subsection{Relationship to existing methods}

The Python baseline is a faithful reimplementation of \ac{ptfce} \cite{spisak2019}, with $r = 0.992$ between baseline and hybrid scores on synthetic data (Section~\ref{sec:concordance}, Table~\ref{tab:concordance}).
Lower concordance on real brain data ($r \approx 0.87$ between R and Python) reflects numerical differences in lookup-table resolution and \ac{grf} parameter estimation; on IXI, the strict subset property indicates that these differences are conservative, and on UK Biobank the three variants remain qualitatively concordant despite small mask-handling differences in the R implementation.

The hybrid extends \ac{etfce} \cite{chen2026} by replacing its permutation inference with \ac{grf} theory.
The union-find incurs a 3-fold additional computational cost on the phantom and approximately 16-fold on whole-brain data relative to the \ac{ccl}-based baseline (Section~\ref{sec:runtime}, Fig.~\ref{fig:smoothness-runtime}b), but it provides exact cluster sizes at every threshold and enables a five-fold denser grid ($n = 500$ versus $100$) that substantially reduces the grid-dependent approximation in the accumulated evidence.

The FSL \ac{tfce} scaling bug identified by Chen et al.\ (the step size $\Delta\tau$ is hard-coded to the statistic-map range rather than treated as a Riemann sum weight, as verified in FSL v6.0.7.19 \cite{jenkinson2012fsl}) is a cautionary example: a subtle numerical error persisted undetected for over fifteen years in one of the most widely used neuroimaging packages, which underscores the sensitivity of discretised integral approximations to implementation details.
The union-find architecture is immune to this class of errors because cluster sizes are retrieved directly from the merge tree rather than recomputed by thresholding; this robustness, combined with the ability to use arbitrarily dense threshold grids at linear cost, justifies the modest additional computational cost.
The grid-convergence analysis (Section~\ref{sec:grid-convergence}, Supplementary Fig.~\ref{fig:supp-grid-convergence}) indicates that the residual approximation is small at the default grid density ($n = 500$, $r > 0.998$, Dice $= 1.0$) and that inference results are stable for $n \geq 200$.
The identity between hybrid and baseline metrics at matched grid density ($n = 100$) further supports the conclusion that the union-find and \ac{ccl} retrieve the same cluster sizes, so any difference between the two methods is attributable solely to grid density.

\subsection{Limitations}

First, the analytical $p$-values assume a stationary \ac{grf} with uniform smoothness, which is violated near grey--white matter boundaries and in regions with heterogeneous cytoarchitecture.
Non-stationary extensions exist \cite{worsley1996} but are not yet incorporated.

Second, the implementation operates on three-dimensional volumes with 26-connectivity; extension to surface-based analyses would require a geodesic union-find on triangulated meshes.

Third, generalisation beyond structural \ac{vbm} to functional MRI (where temporal autocorrelation complicates \ac{grf} assumptions \cite{eklund2016}), diffusion tensor imaging, or arterial spin labelling remains untested.

Fourth, the pure-Python union-find makes the hybrid approximately 16-fold slower than the \ac{ccl}-based baseline on whole-brain data (3-fold on the $64^3$ phantom); a C or Cython extension would close this gap.

Finally, the \ac{grf} lookup table must be computed on first use for each mask size (10--30 seconds); subsequent runs with the same mask reuse the cache.

\section{Conclusion}
\label{sec:conclusion}

This work presented a hybrid \ac{etfce}--\ac{grf} method that combines union-find-based exact cluster-size retrieval with \ac{ptfce}'s analytical \ac{grf} inference and achieves both properties simultaneously for the first time.
A six-experiment Monte Carlo validation indicated that the method controls \ac{fwer} at the nominal level (0/200 null rejections, 95\% CI $[0.0\%, 1.9\%]$), matches the statistical power of baseline \ac{ptfce} (Dice $\geq 0.999$ at sufficient signal strength), and achieves high cross-variant concordance ($r = 0.992$).
The \texttt{pytfce} baseline reimplementation is $75\times$ faster than the reference R~\ac{ptfce} on whole-brain data (${\sim}5$\,s versus ${\sim}390$\,s on ${\sim}2$\,M voxels); the hybrid variant, which provides exact cluster-size retrieval at each threshold, completes in ${\sim}85$\,s ($4.6\times$ faster than R).
Both analytical variants are more than three orders of magnitude faster than permutation-based \ac{tfce}.

On real brain data from UK Biobank ($N = 500$) and IXI ($N = 563$), the method detected biologically plausible scanner, age, and sex effects.
On IXI, the significance maps formed strict subsets of those produced by the R reference, which supports conservative error control; on UK Biobank, the three variants produced qualitatively concordant maps.

The method is implemented in \texttt{pytfce}, an open-source, pip-installable Python package with fewer than 1{,}000 lines of core code and no dependencies on R or FSL \cite{yin2026}.
By reducing \ac{ptfce} runtime by up to two orders of magnitude, the package enables routine use of enhanced \ac{tfce} inference for large-cohort voxelwise neuroimaging analysis that requires both sensitivity and rigorous multiple-comparison correction.

\section*{Data and code availability}

The \texttt{pytfce} package is available at \url{https://github.com/Don-Yin/pytfce} and on PyPI (\texttt{pip install pytfce}). A companion software paper has been submitted to the Journal of Open Source Software.

\section*{Acknowledgements}

We thank Tam\'as Spis\'ak for the original R~\acs{ptfce} implementation.
We acknowledge the use of the IXI dataset (funded by EPSRC GR/S21533/02) and UK Biobank (application 1188243).

\section*{Declaration of competing interest}

The authors declare no competing interests.

\section*{CRediT authorship contribution statement}

\textbf{Don Yin}: Conceptualization, Methodology, Software, Validation, Formal analysis, Investigation, Data curation, Writing -- original draft.
\textbf{Hao Chen}: Resources, Methodology, Software, Formal analysis, Writing -- review \& editing.
\textbf{Takeshi Miki}: Investigation, Writing -- review \& editing.
\textbf{Enyu Yang}: Investigation, Writing -- review \& editing.

\section*{Declaration of generative AI and AI-assisted technologies in the writing process}

During the preparation of this work, the authors used AI-assisted tools for grammar and style checking. The authors reviewed and edited all output and take full responsibility for the content of the publication.

\clearpage
\appendix
\setcounter{figure}{0}
\renewcommand{\thefigure}{S\arabic{figure}}

\section*{Supplementary Material}

\begin{figure*}[htbp]
\centering
\includegraphics[width=0.48\textwidth]{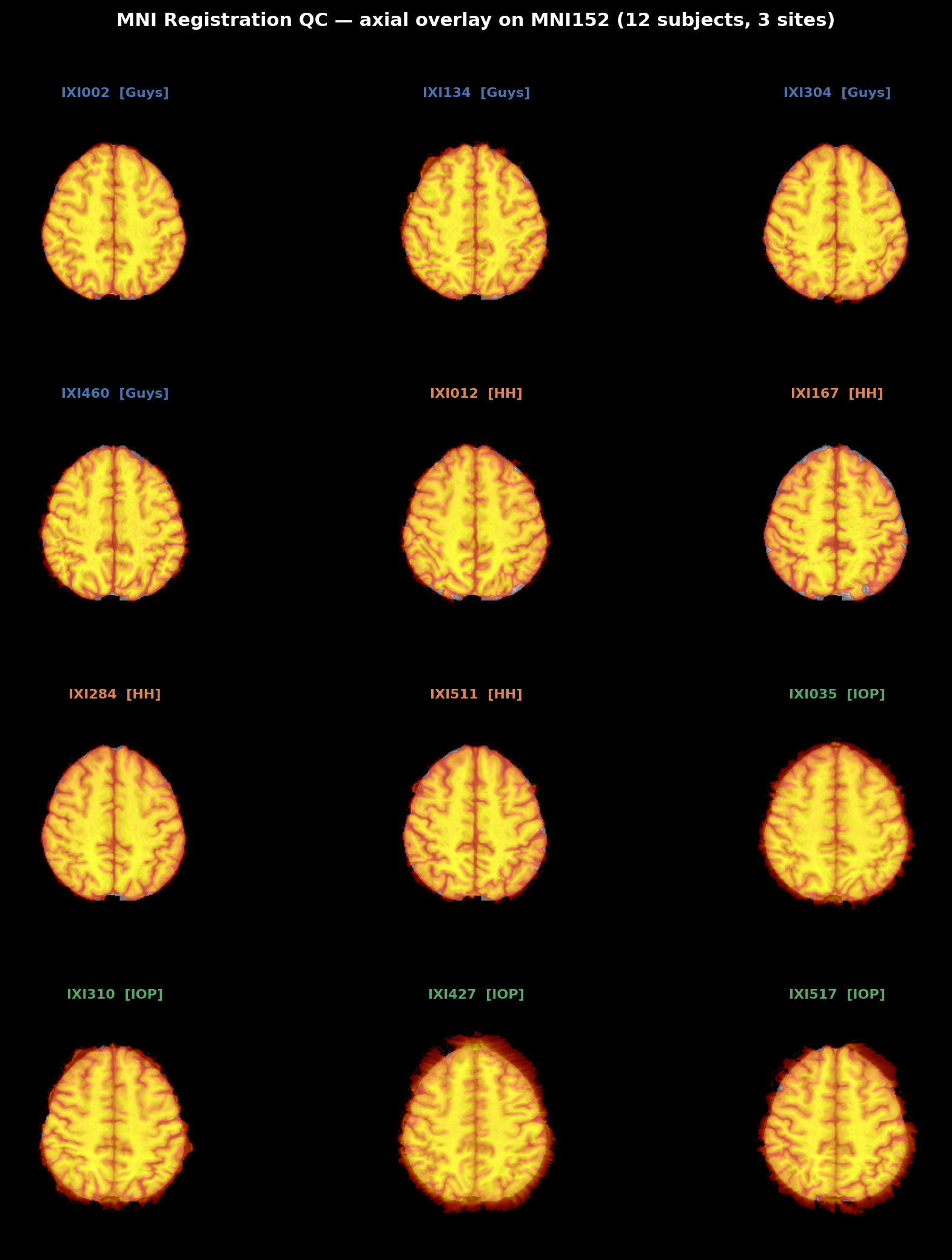}
\hfill
\includegraphics[width=0.48\textwidth]{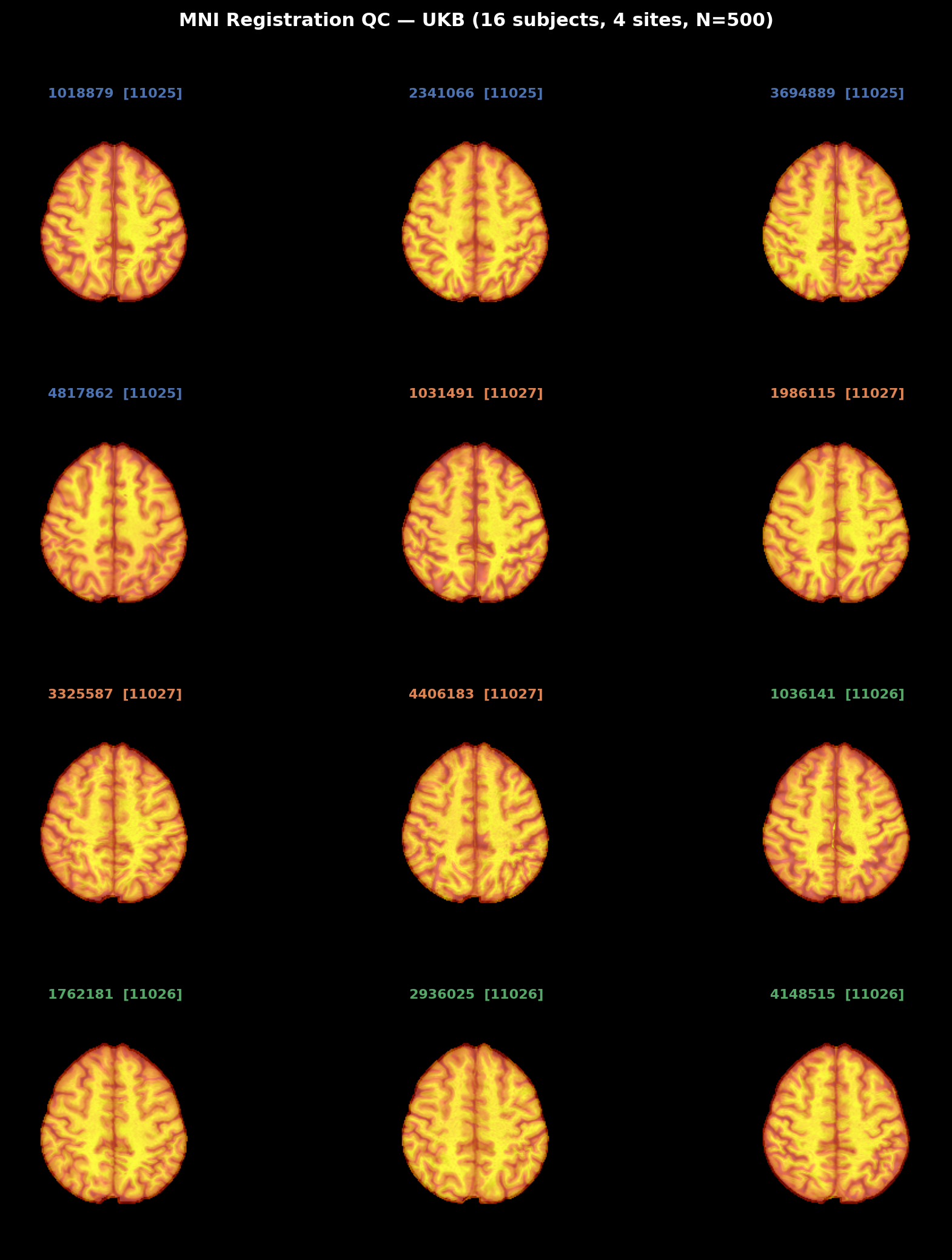}
\caption{Registration quality for IXI (left, 12 representative subjects) and UK Biobank (right). All subjects in both datasets passed visual \acs{qc} after FNIRT nonlinear registration to \acs{mni}152 space.}
\label{fig:supp-registration}
\end{figure*}

\begin{figure*}[htbp]
\centering
\includegraphics[width=0.48\textwidth]{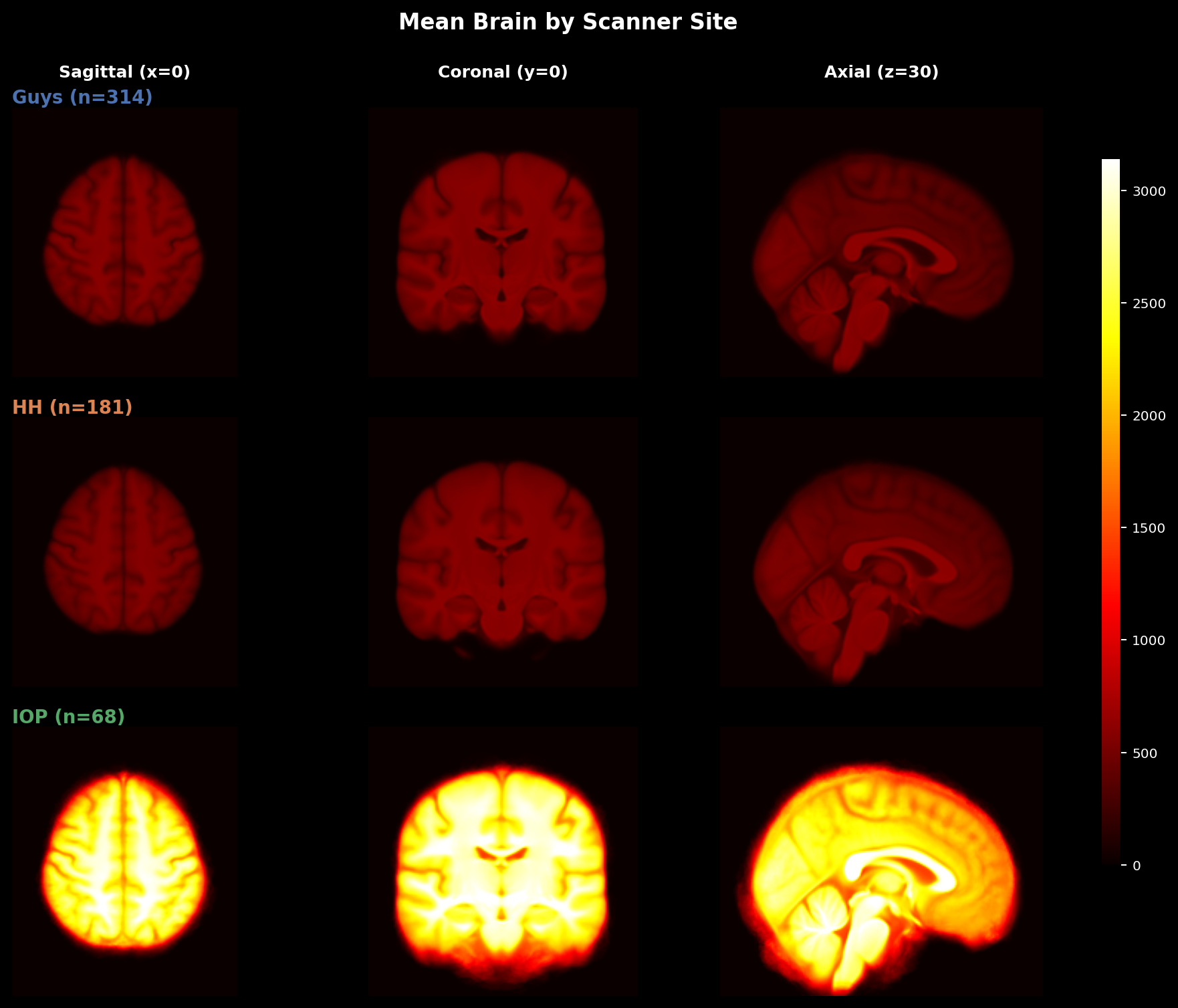}
\hfill
\includegraphics[width=0.48\textwidth]{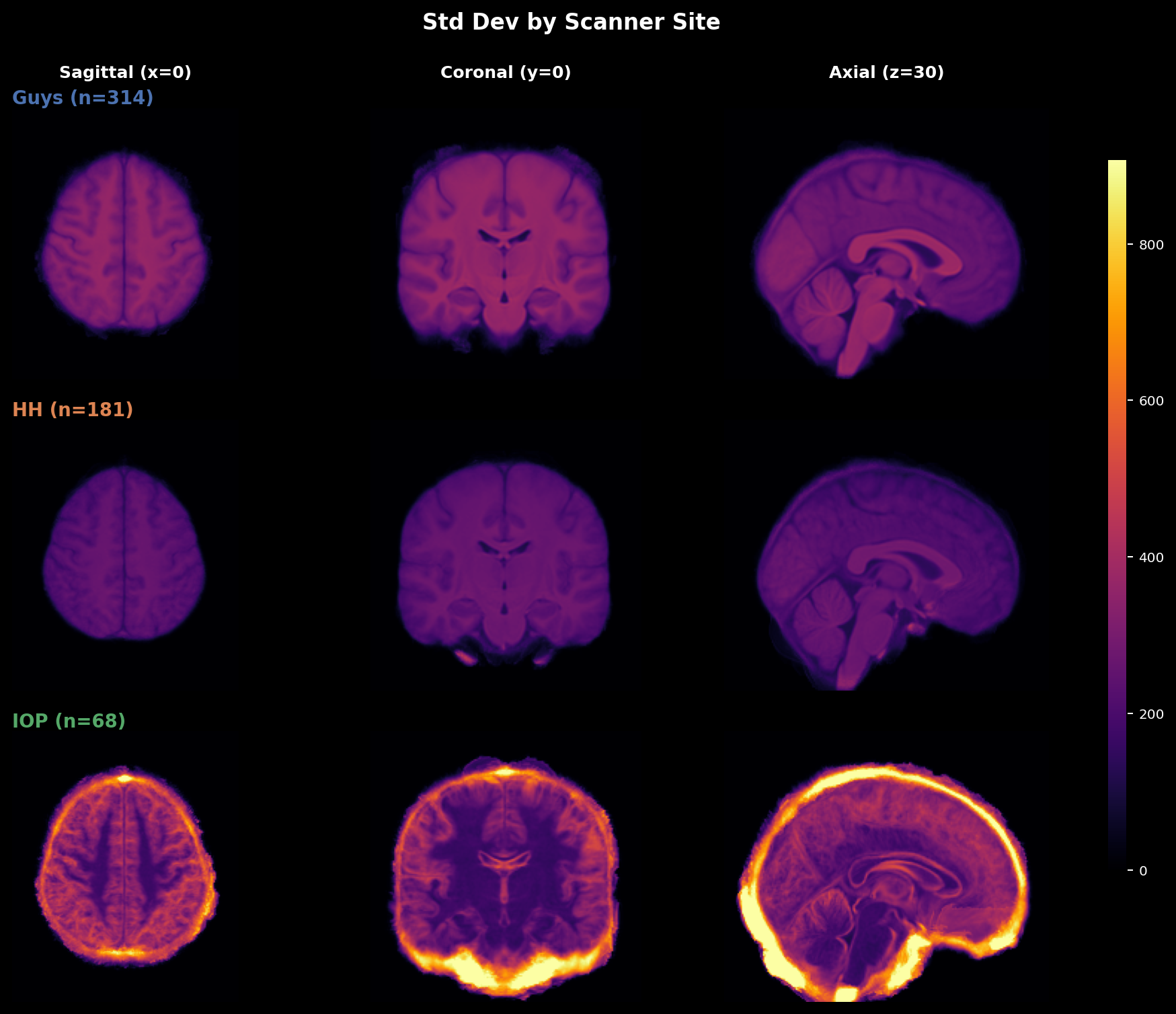}
\caption{IXI site-level mean (left) and standard deviation (right) of registered T1-weighted intensity maps. The IOP site (GE 1.5T) shows higher mean intensity and inter-subject variability compared to the two Philips sites, which reflects cross-vendor scanner differences that the voxelwise analysis is designed to detect.}
\label{fig:supp-ixi-site}
\end{figure*}

\begin{figure*}[htbp]
\centering
\includegraphics[width=0.48\textwidth]{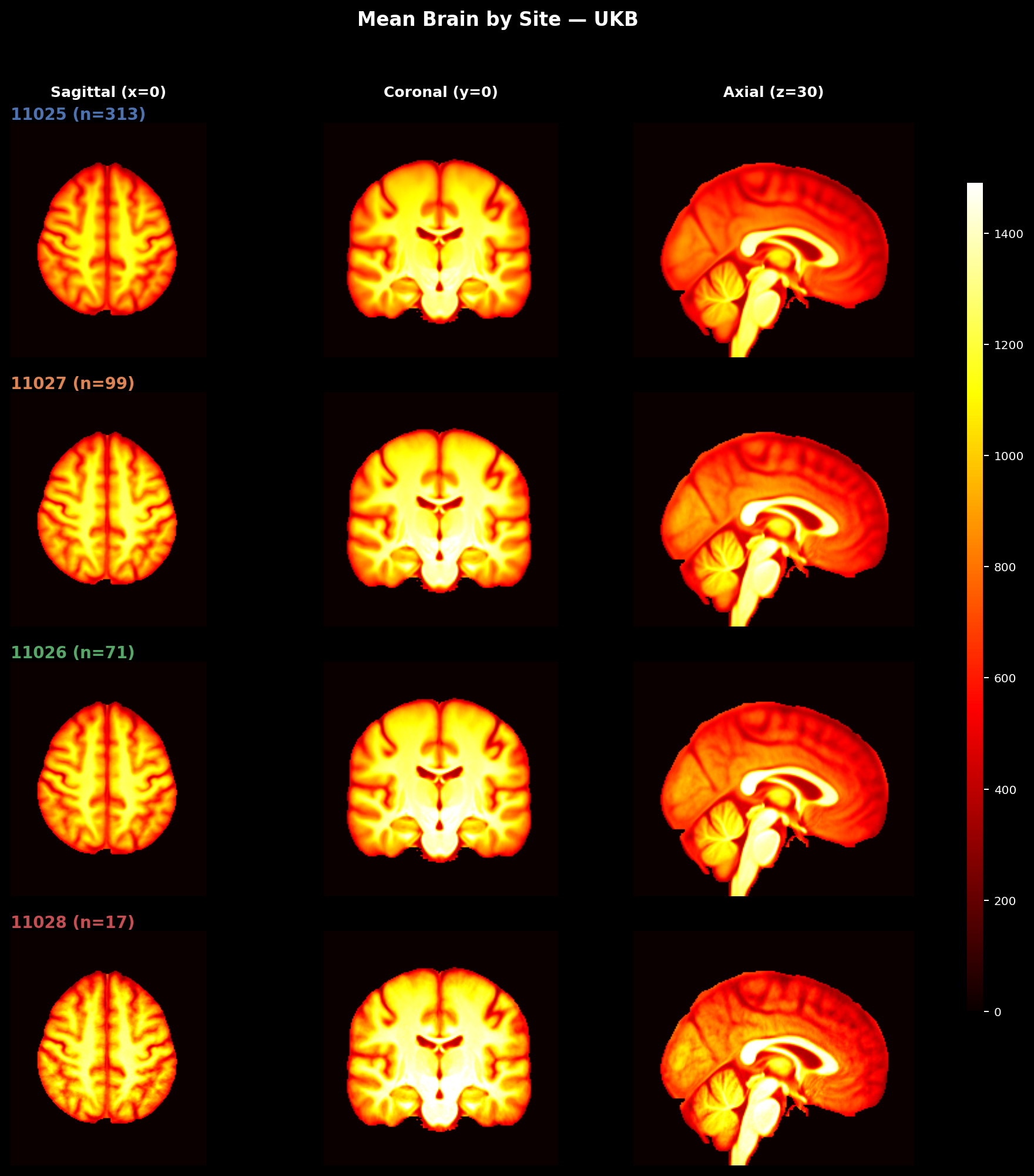}
\hfill
\includegraphics[width=0.48\textwidth]{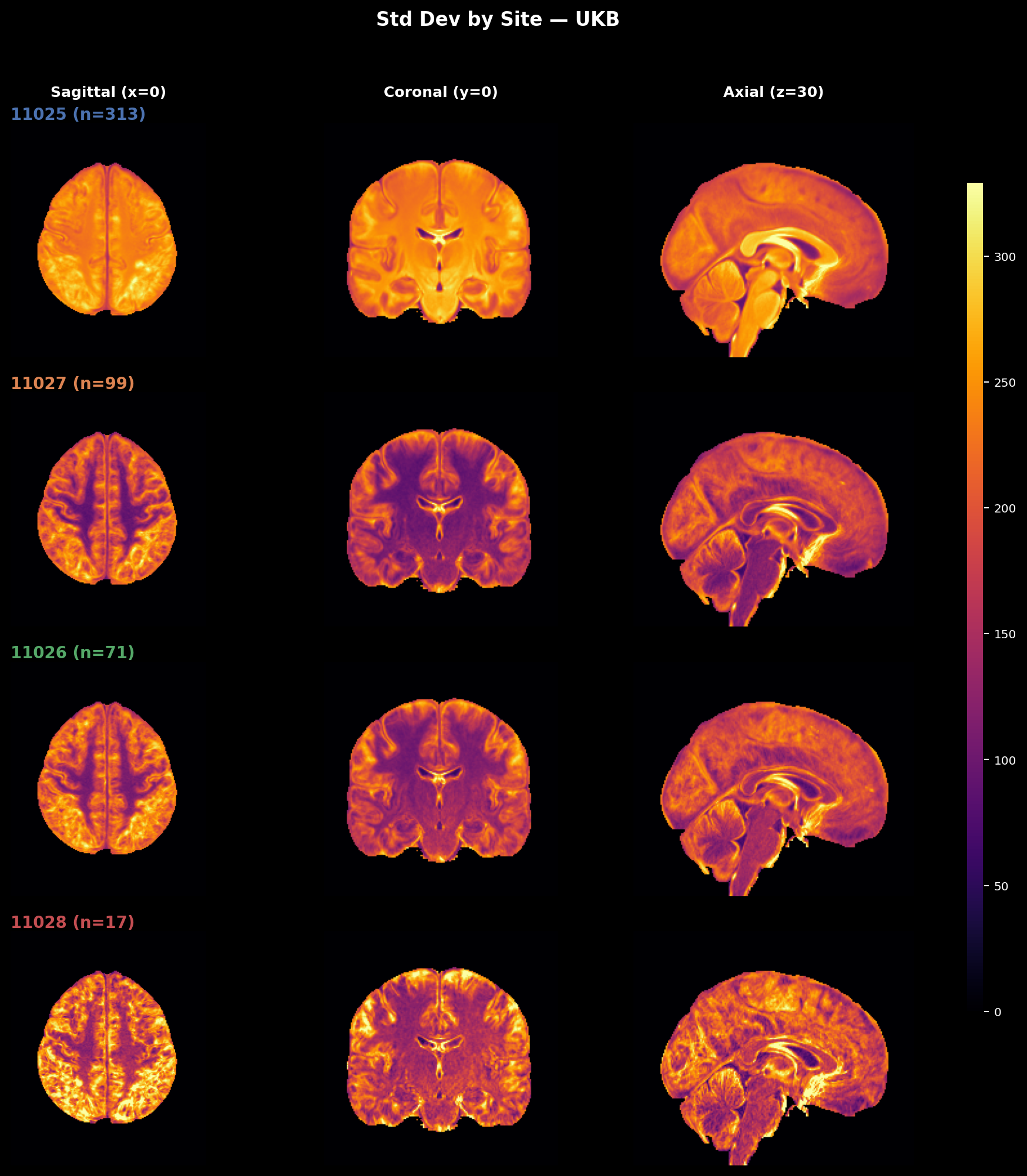}
\caption{UK Biobank site-level mean (left) and standard deviation (right) of registered T1-weighted intensity maps. Within-vendor variability across the four Siemens Skyra sites is substantially smaller than the cross-vendor variability observed in IXI (Fig.~\ref{fig:supp-ixi-site}).}
\label{fig:supp-ukb-site}
\end{figure*}

\begin{figure}[htbp]
\centering
\includegraphics[width=\columnwidth]{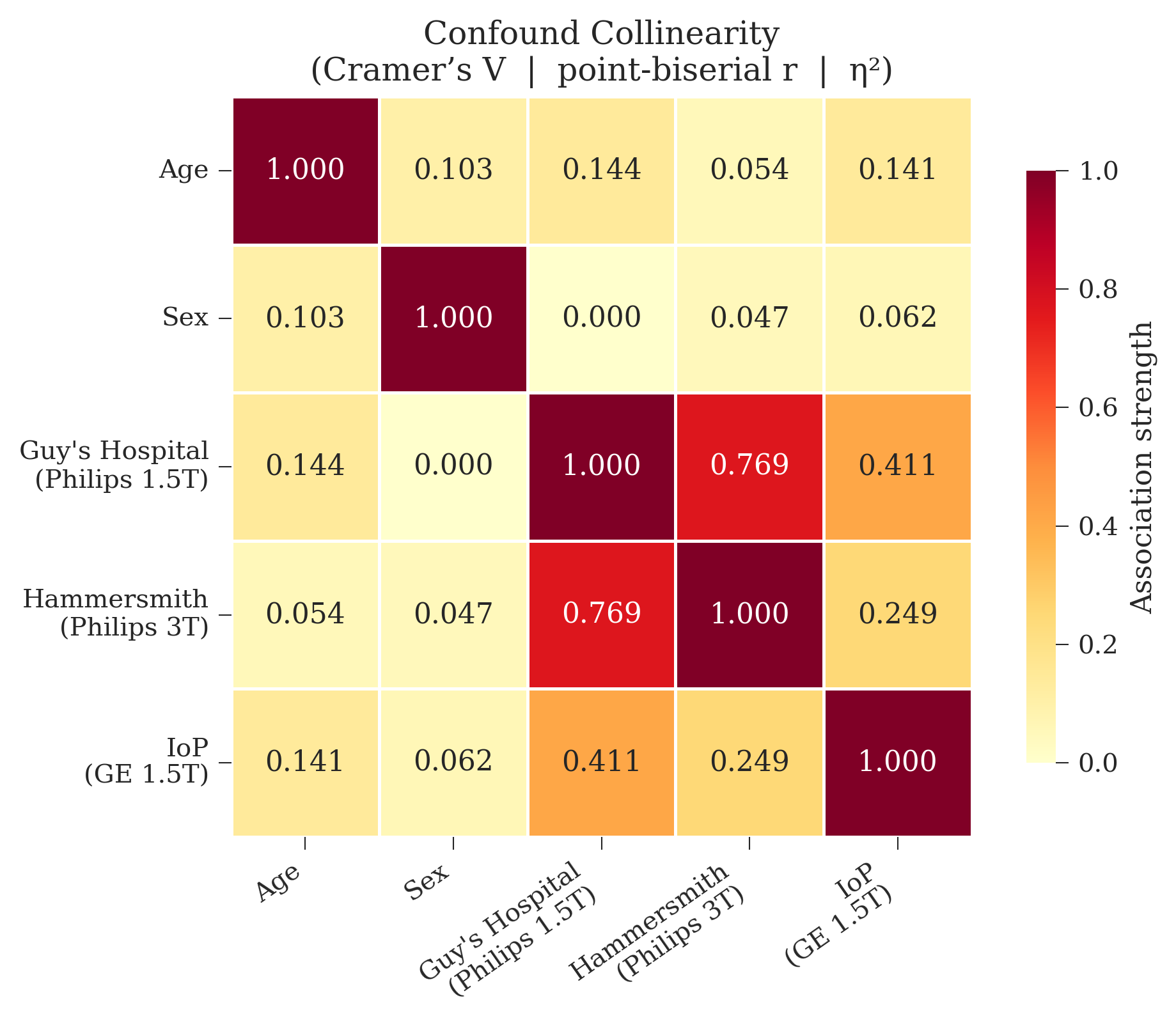}
\caption{Confound collinearity heatmap for the IXI dataset. Age--site association ($\eta^2 \leq 0.14$) and sex--site association ($|r| \leq 0.06$) are modest, which supports the interpretability of site-effect maps. Inter-site Cram\'er's $V$ (0.41--0.77) reflects the unequal site sample sizes rather than demographic confounding.}
\label{fig:supp-confound}
\end{figure}

\begin{figure}[htbp]
\centering
\includegraphics[width=\columnwidth]{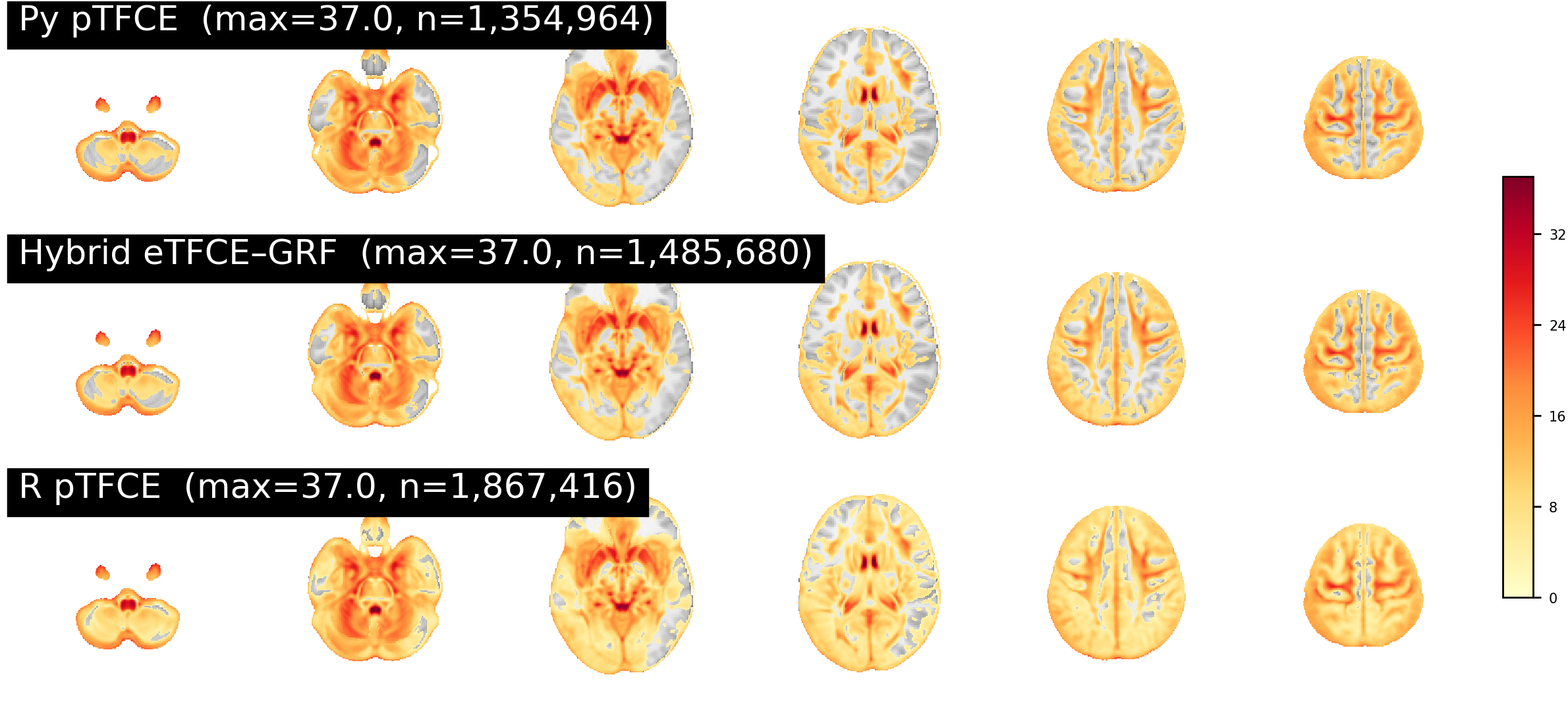}\\[6pt]
\includegraphics[width=\columnwidth]{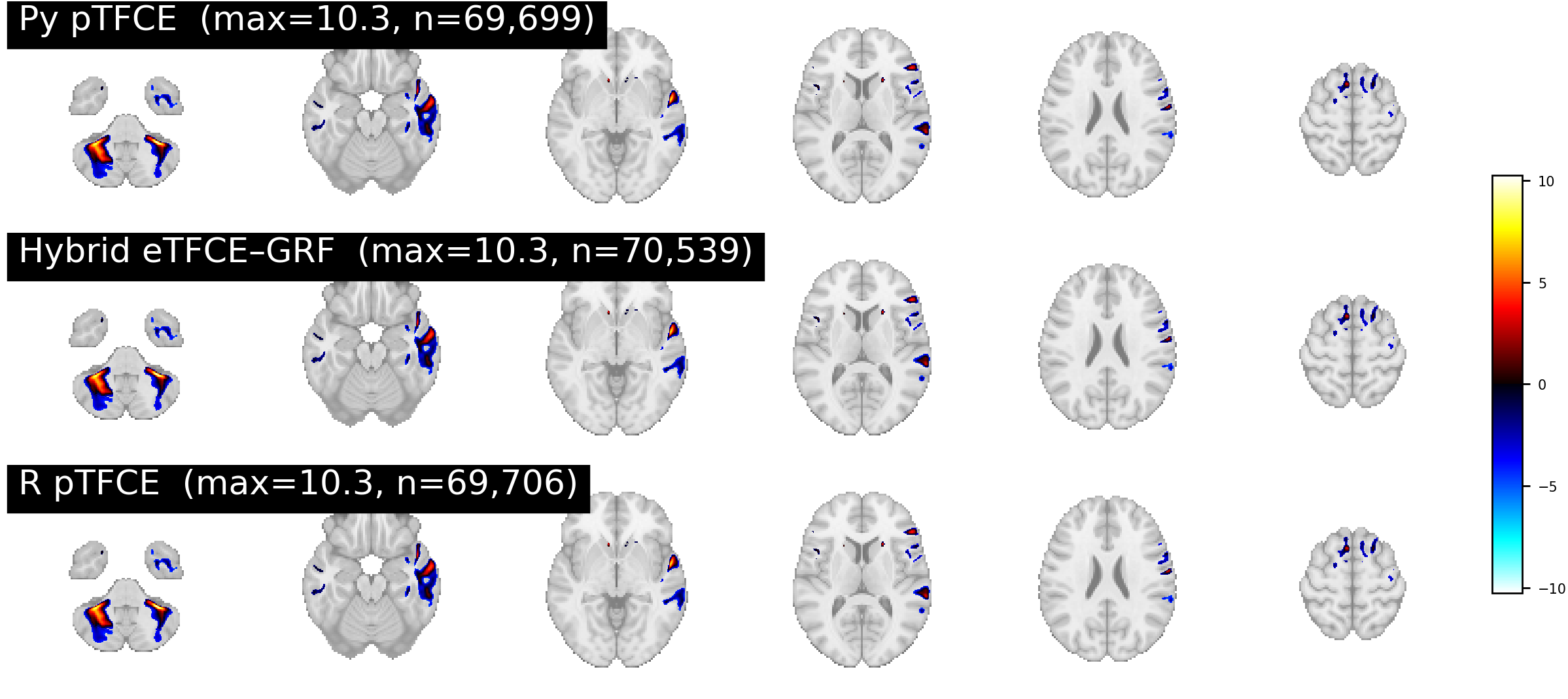}\\[6pt]
\includegraphics[width=\columnwidth]{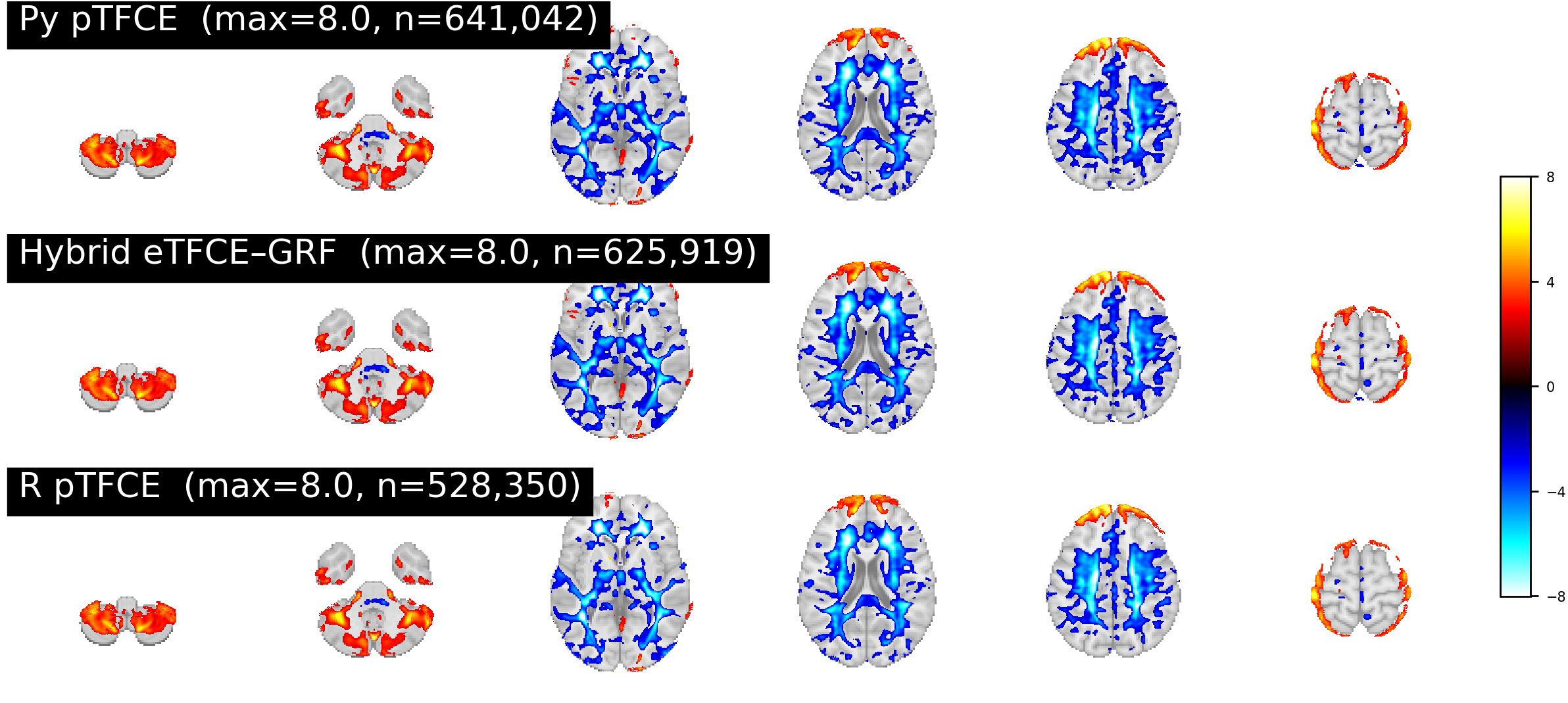}
\caption{IXI per-variant significance maps.
(a)~Site effect, (b)~age effect, (c)~sex effect.
Each panel shows raw statistics masked to \ac{fwer} $p < 0.05$ for the three \ac{ptfce} variants (Python baseline, hybrid eTFCE--GRF, R reference). All three variants produce qualitatively identical spatial patterns.}
\label{fig:supp-ixi-variants}
\end{figure}

\begin{figure}[htbp]
\centering
\includegraphics[width=\columnwidth]{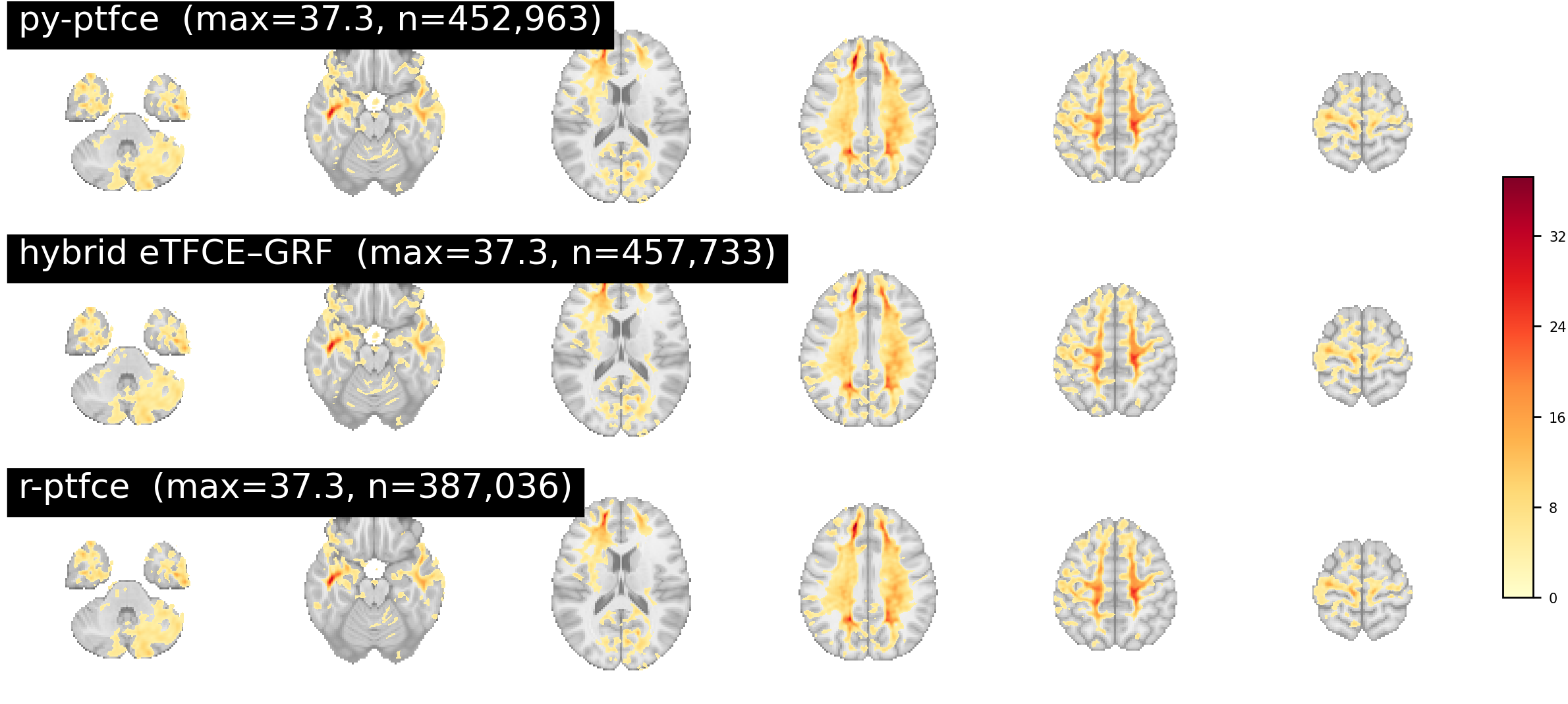}\\[6pt]
\includegraphics[width=\columnwidth]{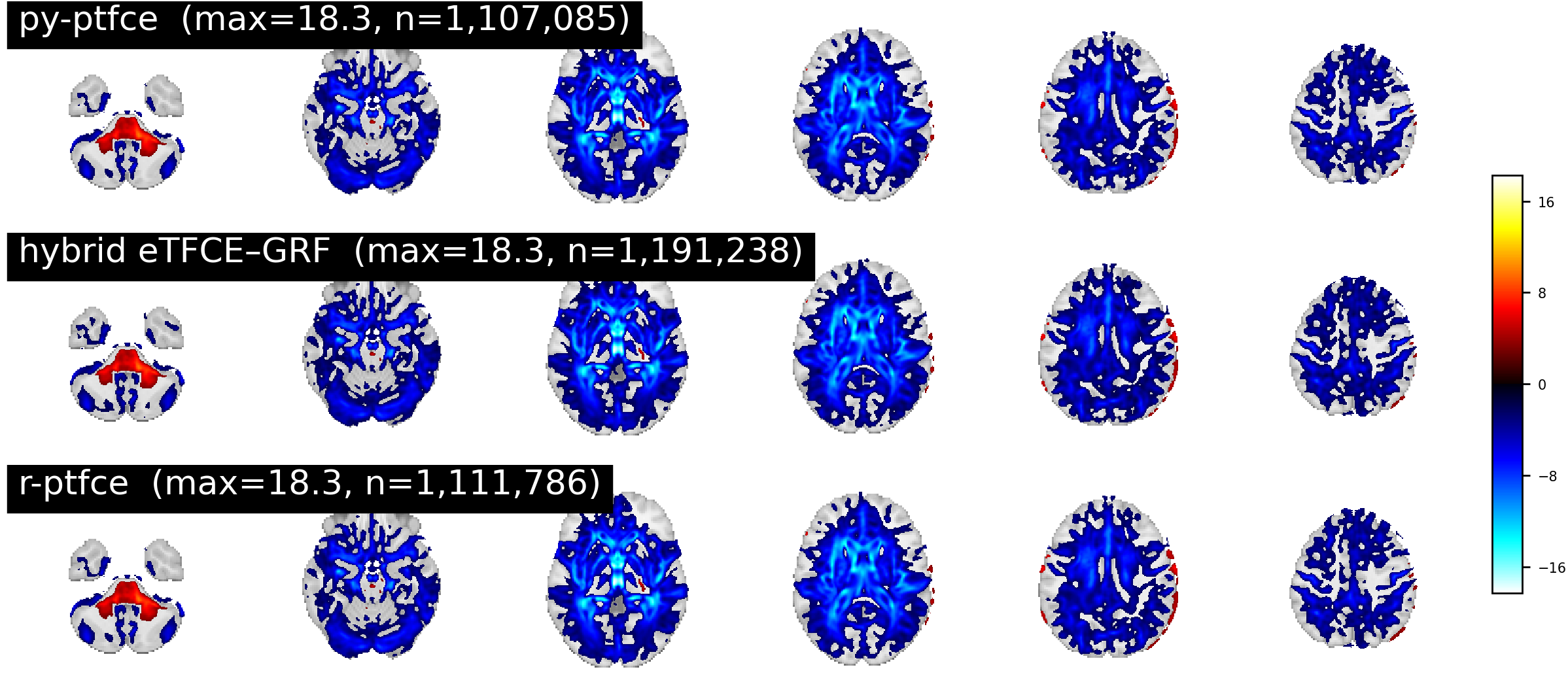}\\[6pt]
\includegraphics[width=\columnwidth]{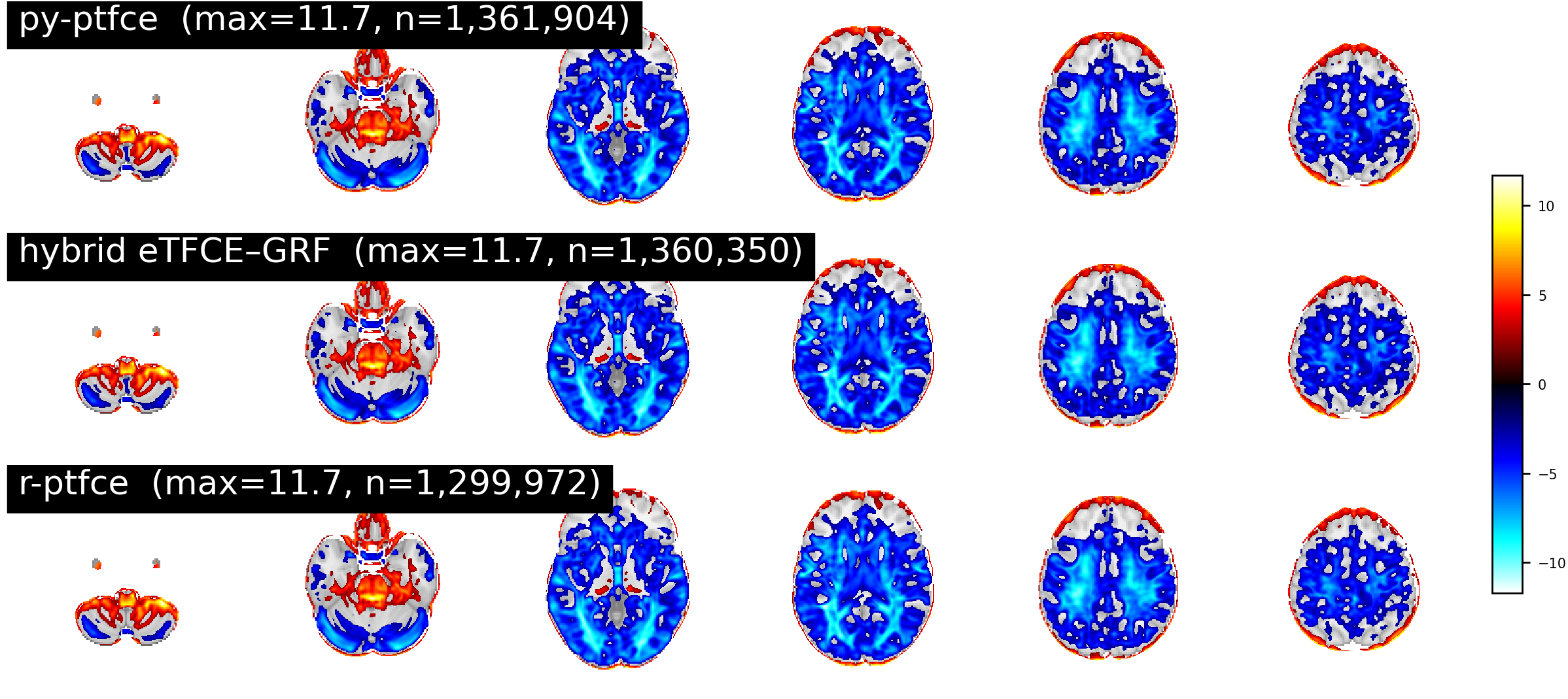}
\caption{UK Biobank per-variant significance maps.
(a)~Scanner effect, (b)~age effect, (c)~sex effect.
As in the IXI analysis (Fig.~\ref{fig:supp-ixi-variants}), the three \ac{ptfce} variants produce consistent spatial patterns across all contrasts.}
\label{fig:supp-ukb-variants}
\end{figure}

\begin{figure}[htbp]
\centering
\includegraphics[width=\columnwidth]{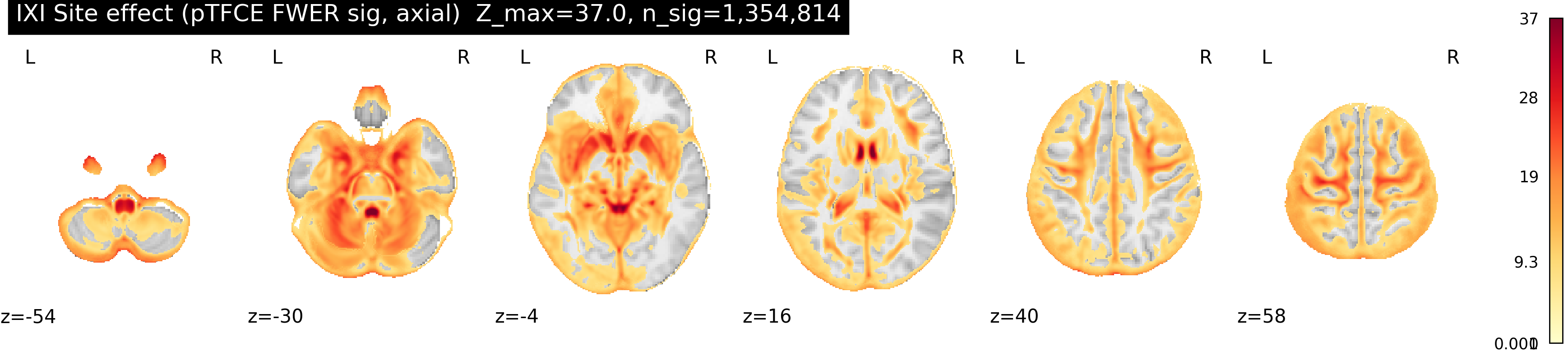}\\[4pt]
\includegraphics[width=\columnwidth]{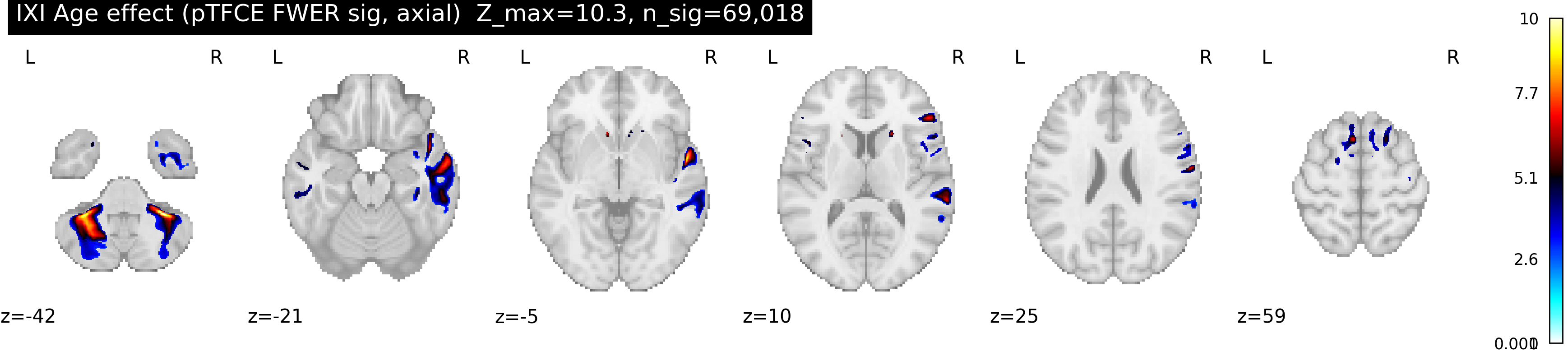}\\[4pt]
\includegraphics[width=\columnwidth]{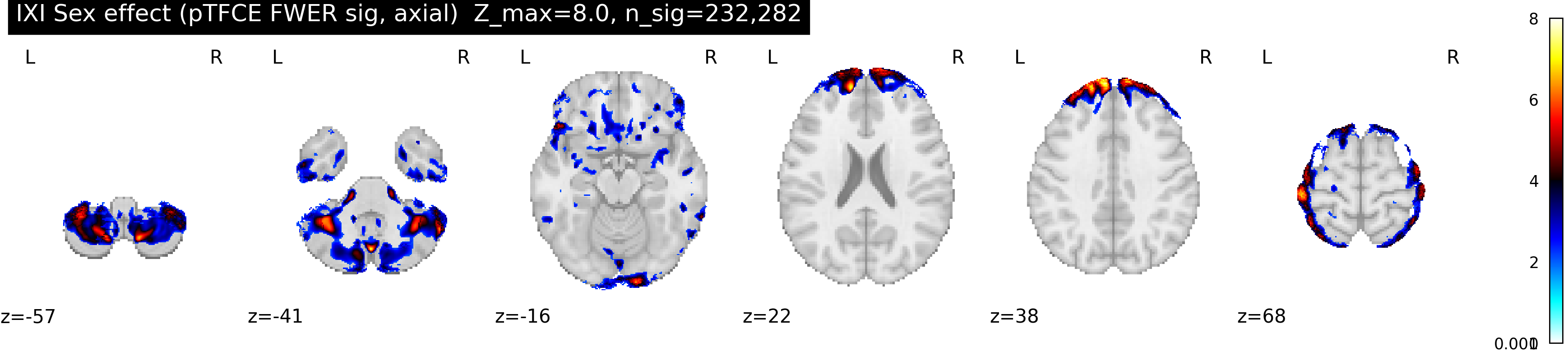}\\[8pt]
\includegraphics[width=\columnwidth]{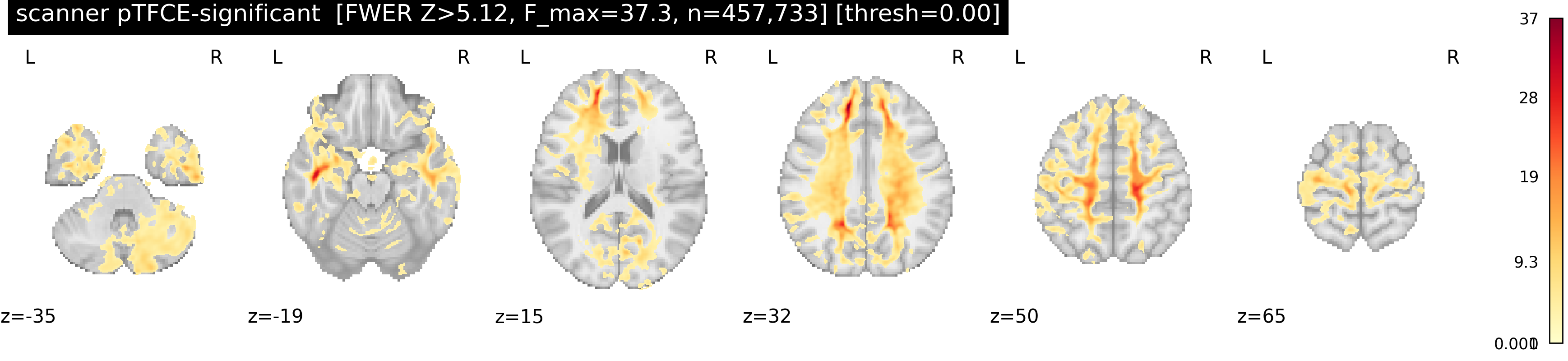}\\[4pt]
\includegraphics[width=\columnwidth]{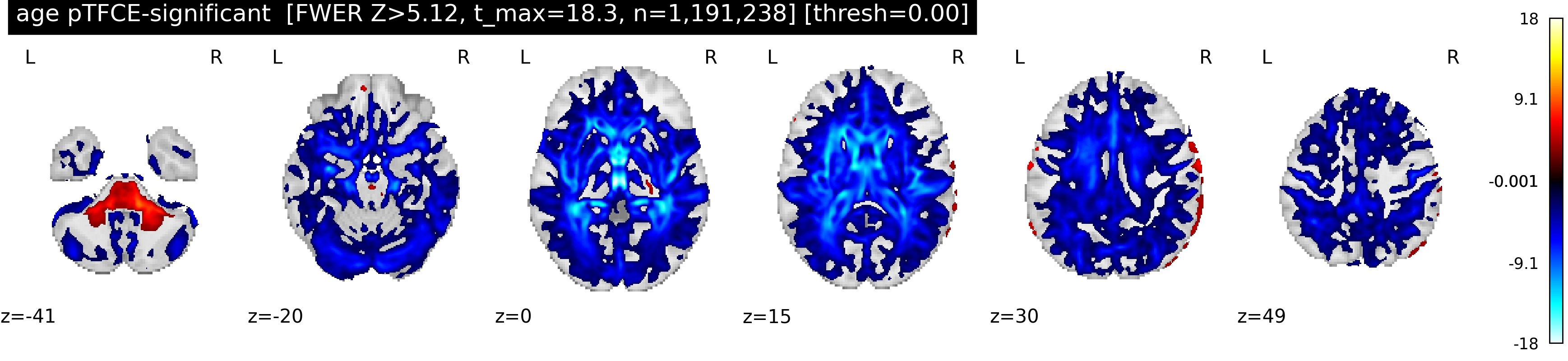}\\[4pt]
\includegraphics[width=\columnwidth]{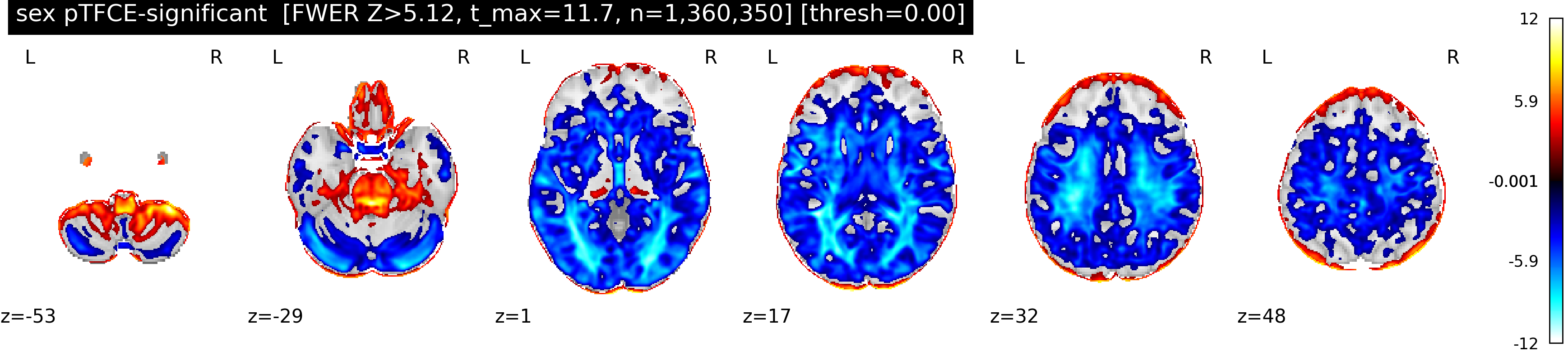}
\caption{Axial slice views of significance maps for all six contrasts.
(a)~IXI site effect, (b)~IXI age effect, (c)~IXI sex effect,
(d)~UK Biobank scanner effect, (e)~UK Biobank age effect, (f)~UK Biobank sex effect.
All maps are thresholded at \ac{fwer} $p < 0.05$ (Python \ac{ptfce} baseline).}
\label{fig:supp-axial}
\end{figure}

\begin{figure}[htbp]
\centering
\includegraphics[width=\columnwidth]{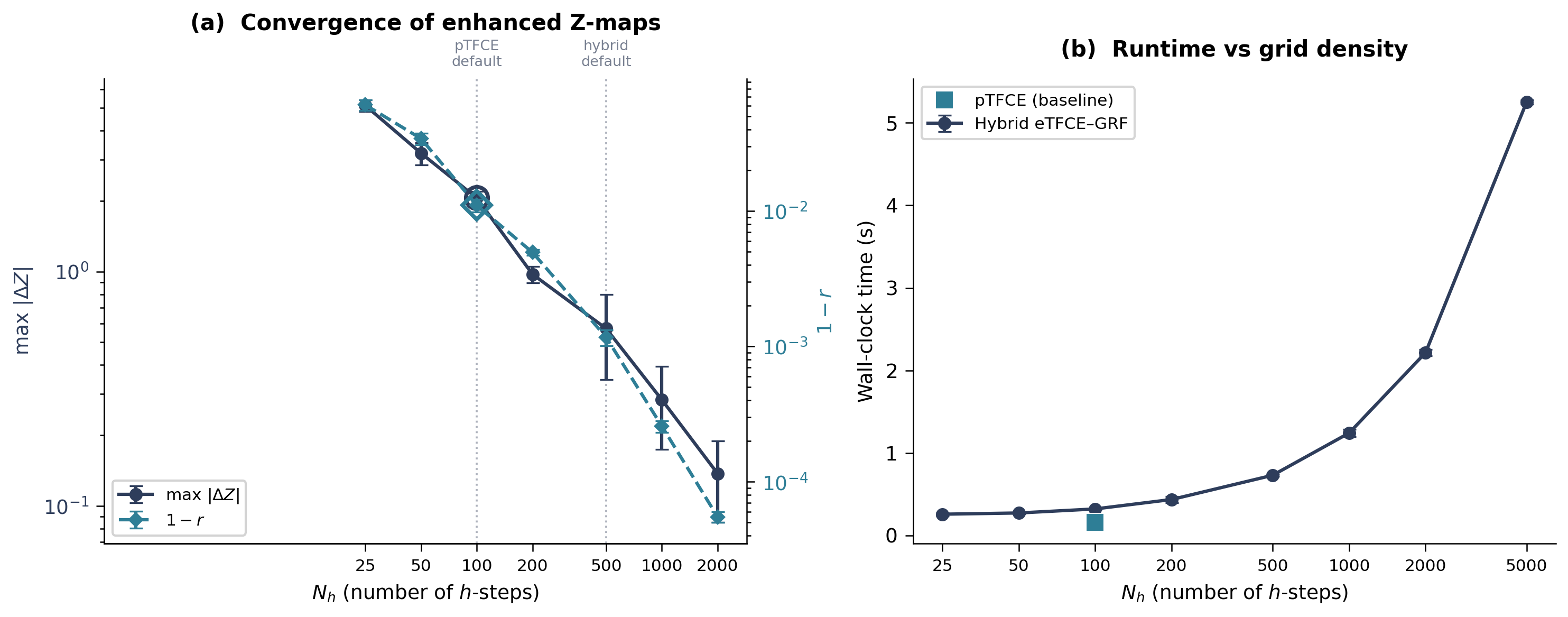}
\caption{Convergence of hybrid eTFCE--GRF enhanced $Z$-maps with increasing threshold grid density $n$.
(a)~Maximum absolute difference $\max|\Delta Z|$ (left axis, navy) and $1 - r$ (right axis, teal) relative to a converged reference ($n = 5000$), as a function of $n$ on a log scale.
Vertical dashed lines mark the baseline \acs{ptfce} default ($n = 100$) and the hybrid default ($n = 500$).
At $n = 500$, $\max|\Delta Z| = 0.57 \pm 0.23$ (mean $\pm$ SD) and $r > 0.998$; the Dice coefficient of significant voxel sets exceeds $0.999$ by $n = 200$ and reaches $1.0$ by $n = 500$.
Open markers at $n = 100$ show the baseline \acs{ptfce} result, which produces identical metrics to the hybrid at matched grid density.
Error bars denote $\pm 1$ SD across three independent phantom realisations.
(b)~Wall-clock time scales approximately linearly with $n$ beyond the fixed union-find construction cost, which indicates that denser grids are affordable.}
\label{fig:supp-grid-convergence}
\end{figure}

\bibliographystyle{elsarticle-num}
\bibliography{main}

\end{document}